\documentstyle[eqsecnum,aps,prd,epsfig]{revtex}
\date{\today}
\tighten
\begin{document}
\tighten

%
\let\ov=\over
\let\lbar=\l
\let\l=\left
\let\r=\right
\def \der#1#2{{\partial{#1}\over\partial{#2}}}
\def \dder#1#2{{\partial^2{#1}\over\partial{#2}^2}}
\def\N{{I\!\!N}}
\def\be{\begin{equation}}
\def\ee{\end{equation}}
\def\beu{\begin{displaymath}}
\def\eeu{\end{displaymath}}
\def\bea{\begin{eqnarray}}
\def\eea{\end{eqnarray}}
\def\beau{\begin{eqnarray*}}
\def\eeau{\end{eqnarray*}}
\def\ms{\langle S \rangle}
\def\n2{\langle N^2 \rangle}
\def\sn2{\sqrt{\langle N^2 \rangle}}
\def\sech{\mathop{\rm sech}\nolimits}

%

%
\title{Detection of Anisotropies in the Gravitational-Wave Stochastic
Background}
\author{Bruce Allen}
\address{Department of Physics\\
University of Wisconsin - Milwaukee\\
PO Box 413\\
Milwaukee, WI 53211, USA\\
email: ballen@dirac.phys.uwm.edu}

\author{Adrian C. Ottewill} \address{Department of Mathematical
Physics, University College Dublin,\\ Belfield, Dublin 4, Ireland\\
email: ottewill@relativity.ucd.ie}

\maketitle
\begin{abstract}
By correlating the signals from a pair of gravitational-wave detectors,
one can undertake sensitive searches for a stochastic background of
gravitational radiation.  If the stochastic background is anisotropic,
then this correlated signal varies harmonically with the earth's
rotation.  We calculate how the harmonics of this varying signal are
related to the multipole moments which characterize the anisotropy, and
give a formula for the signal-to-noise ratio of a given harmonic.  The
specific case of the two LIGO (Laser Interferometric Gravitational
Observatory) detectors, which will begin operation around the year
2000, is analyzed in detail.  We consider two possible examples of
anisotropy.  If the gravitational-wave stochastic background contains a
dipole intensity anisotropy whose origin (like that of the Cosmic
Background Radiation) is motion of our local system, then that
anisotropy will be observable by the advanced LIGO detector (with 90\%
confidence in one year of observation) if $\Omega_{\rm gw} > 5.3 \times
10^{-8} h_{100}^{-2} $.  We also study the signal produced by
stochastic sources distributed in the same way as the luminous matter
in the galactic disk, and in the same way as the galactic halo.
The anisotropy due to sources distributed as the galactic disk 
or as the galactic halo will be
observable by the advanced LIGO detector (with 90\%
confidence in one year of observation) if $\Omega_{\rm gw} > 1.8 \times
10^{-10} h_{100}^{-2} $ or  $\Omega_{\rm gw} > 6.7 \times
10^{-8} h_{100}^{-2} $, respectively.
\end{abstract}

\section{INTRODUCTION}
The design and construction of a number of new and more sensitive
detectors of gravitational radiation is currently underway.  These
include the LIGO detector being built in the United States by a joint
Caltech/MIT collaboration \cite{science92}, the VIRGO detector being
built near Pisa by an Italian/French collaboration \cite{virgo}, the
GEO-600 detector being built in Hannover by an Anglo/German
collaboration \cite{geo600}, and the TAMA-300 detector being built near
Tokyo \cite{tama300}.  There are also several resonant bar detectors
currently in operation, and several more refined bar and
interferometric detectors presently in the planning and proposal
stages.

When two or more of these detectors are operating, it will become
possible to correlate their signals, and in this way, to search for a
stochastic background of gravitational radiation.  The technique for
such a search was originally described in work by Michelson
\cite{mich}, Cristensen \cite{chris} and Flanagan \cite{flan}.  A
review of these techniques may be found in \cite{myreview}.  Such
radiation might be the result of processes that took place during the
very early universe.  It might also result from the incoherent
superposition of many faint unresolvable present-day sources such as
coalescing binary systems.

The stochastic gravitational-wave background might be isotropic on the
sky, or it might be anisotropic.  For example, if the background
results from early-universe processes, then it might be isotropic to
about the same degree as the $2.7^\circ\> \rm K$ electromagnetic
background radiation.  On the other hand, if the background is due to
white-dwarf binaries in our own galaxy, then they might be distributed
in a pancake or bar which mimics the shape of the observed luminous
matter in our galaxy.  In this paper, we show how the correlated signal
from a pair of gravitational wave detectors is related to multipole
moments which characterize the anisotropy.   This should permit a
signal to be analyzed to search for (or place upper limits on) the
multipole moments which characterize the anisotropy.  In this paper, we
will assume that the reader is already familiar with the work
previously cited (references \cite{mich,chris,flan,myreview}) on
stochastic background detection.

These paper is organized as follows.  In Section~\ref{s:first} we show
how a background of stochastic gravitational radiation may be
decomposed in a plane-wave expansion, with the coefficients of the
expansion treated as stochastic random variables. In
Section~\ref{s:spect} the properties of these random variables are
related to the (frequency) spectrum and spatial distribution of the
radiation, and a set of multipole moments are introduced which
characterize the anisotropies of the stochastic background.  These
anisotropies may be searched for by studying the variations of the
detector outputs as the earth rotates relative to the fixed cosmic
frame.  In Section~\ref{s:howto} we show how the correlation between a
pair of detectors fixed on the earth varies with time as the earth
rotates, and detail how that correlation is related to the anisotropies
of the stochastic gravitational background.  The variation of the
correlation with the earth's rotation may be decomposed into harmonics
of the earth's period.  In section Section~\ref{s:rothar} we introduce
a set of functions $\gamma_{\ell m}(f)$ which are generalizations of
the well-known overlap reduction function $\gamma(f)$ of references
\cite{mich,chris,flan,myreview}.  These functions characterize the
effect of the $\ell,m$ anisotropy multipole on the $m$'th harmonic of
the detector correlation.  The principal result of this paper is to
compute these functions for the LIGO pair of detectors.  This is done
by introducing two special frames of reference, fixed with respect to
the earth, in Section~\ref{s:frames}, and then performing a set of
integrations in Section~\ref{s:ccomp}.  The final integrations are
performed and explicit formula for the $\gamma_{\ell m}(f)$ are
obtained in Section~\ref{s:integ}.  In Section~\ref{s:sigtonoise} we
analyze the signal-to-noise ratios associated with the $m$'th harmonic,
and give a formula which may be used to determine if a given anisotropy
is detectable or not.  Following this, we consider two specific
examples of anisotropy.  In Section~\ref{s:ani} we consider a dipole
anisotropy in the stochastic background resulting from our local proper
motion.  In order to predict the harmonics which result, it is
necessary to adopt conventions for the normalization of the optimal
filters; these choices are detailed in Section~\ref{s:norm}.  In
Section~\ref{s:galactic} we then consider the anisotropies in the
stochastic gravitational wave background that would arise from sources
distributed in the same way as our galaxy, and it's halo. This is
followed by a short conclusion.

Throughout this paper, $c$ denotes the speed of light and $G$ denotes
Newton's gravitational constant.

\section{The Stochastic Background}
\label{s:first}
The gravitational wave background may be described in terms of a
perturbation to the Minkowski metric of space-time:
\be
ds^2 = -c^2 dt^2 + d\vec x^2 + h_{ab}(t,\vec x) dx^a dx^b.
\ee
In transverse
traceless gauge, this can be written in the form of a plane wave
expansion as
\be \label{e:plane}
h_{ab}(t,\vec x) =  \sum_A \int_{-\infty}^\infty df \int_{S^2} d\hat
\Omega \> h_A(f,\hat \Omega) {\rm e}^{2 \pi i f(t-\hat \Omega \cdot \vec x/c)}
e_{ab}^A(\hat \Omega).
\ee
Here $h_A(f,\hat \Omega)$ is an arbitrary complex function satisfying
the relation $h_A(-f,\hat \Omega) = h_A^*(f,\hat \Omega)$.  The
polarization states are labeled by $A=+,\times$ and $\hat \Omega$ is a
unit vector on the two-sphere.  The wave-vector of the
corresponding component of the perturbation is
$\vec k = 2 \pi f \hat \Omega/c$.  The polarization tensors $e_{ab}^A$
appearing in these relations may be given explicitly.   In standard
angular coordinates $(\theta,\phi)$ on the two-sphere one may write
\bea
\hat \Omega & = & \cos \phi \sin \theta \hat x + \sin \phi \sin \theta \hat y
+ \cos \theta \hat z \\
\label{e:mdef}
\hat m& = & \sin \phi \hat x - \cos \phi \hat y \\
\label{e:ndef}
\hat n& = & \cos \phi \cos \theta \hat x + \sin \phi \cos \theta \hat y 
- \sin \theta \hat z
\eea
and then choose
\bea
\label{e:polar}
e_{ab}^+(\hat \Omega) & = & m_a m_b - n_a n_b \\
\label{e:polar2}
e_{ab}^\times(\hat \Omega) & = & m_a n_b + n_a m_b
\eea
To simplify matters later (but without any loss of generality) we
assume that the $\hat z$ vector points along the direction of the
earth's rotation axis.  One can verify by inspection that $\hat m$ and
$\hat n$ are a pair of orthogonal unit-length vectors in the plane
perpendicular to $\hat \Omega$.  It is simple to show that any rotation
of the vectors $\hat m$ and $\hat n$ within the plane that they define
simply corresponds to a trivial re-definition of the complex wave
amplitudes $h_+$ and $h_\times$.

To describe a stochastic source, we treat the complex amplitude
$h_A(f,\hat \Omega)$ as a random variable with zero mean value.  In
this paper, we consider stochastic sources which are {\it not
isotropic}. In principle, such a source has spectral properties which
depends upon amplitude and frequency in an arbitrary way.  For
simplicity, in this paper {\it we consider only stochastic sources whose
directional dependence is frequency-independent}. The dependence
of the stochastic background on frequency and direction may be stated
in terms of the expectation value of the product of two random
variables $h_A(f,\hat \Omega)$:
\be \label{e:quad}
\langle h^*_A(f,\hat \Omega) h_{A'}(f',{\hat \Omega}') \rangle =
\delta_{AA'} \delta(f-f') \delta^2(\hat \Omega,{\hat \Omega}') H(f) P(\hat \Omega).
\ee
Here $\delta^2(\hat \Omega,{\hat \Omega}')$ is a covariant
two-dimensional delta-function on the unit two-sphere.  For a general
stochastic source, the quantity $H(f) P(\hat \Omega)$ which appears on
the right hand side would be an arbitrary function of frequency and
direction.  However our assumption that the directional dependence is
frequency-independent implies that the r.h.s. factors as shown. (Note
that the expressions which we later derive for Signal-to-Noise ratios
and expected signal strengths may be trivially extended to include the
most general case.)

\section{Spectrum of the Stochastic Background}
\label{s:spect}
The function $H(f)$ determines the spectrum of the gravitational
radiation.  The energy density in gravitational waves is given by
\be 
\rho_{\rm gw} = {c^2 \over 32 \pi G} \langle \dot h_{ab} \dot h^{ab}
\rangle,
\ee
where the overdot denotes a time derivative, and both tensors are
evaluated at the same space-time point $(t,\vec x)$.
Substituting the plane wave expansion (\ref{e:plane}) into this formula
and using (\ref{e:quad}) yields
\be 
 \langle \dot h_{ab}(t,\vec x) \dot h^{ab}(t,\vec x) \rangle
= \sum_A \int_{-\infty}^\infty df \int_{S^2} d \hat \Omega \>
4 \pi^2 f^2 H(f) P(\hat \Omega) e^A_{ab}(\hat \Omega) e_A^{ab}(\hat \Omega).
\ee
Since $\sum_A e_{ab}^A e^{ab}_A = 4$  
one has
\be \label{e:rhoingw} \langle \dot h_{ab} \dot h^{ab} \rangle=16 \pi^2
\int d\hat \Omega P(\hat \Omega)
 \int_{-\infty}^\infty df \> f^2 H(f) = 32 \pi^2 \int d\hat \Omega
 P(\hat \Omega) \int_0^\infty df \> f^2 H(f).
\ee
In describing gravitational wave stochastic backgrounds, it is
conventional to compare the energy density to the critical energy
density $\rho_{\rm critical}$ required (today) to close the universe.
This critical energy density is determined by the rate at which the
universe is expanding today.  Let us denote the Hubble expansion rate
today by
\be \label{e:hubble}
H_0 = 
100 \> h_{100} \> {{\rm  km}\,{\rm sec}^{-1} \over {\rm Mpc}} = 
3.2 \times 10^{-18} h_{100} \> {\rm  sec}^{-1}  = 
1.1 \times 10^{-28} c \> h_{100} \> {\rm cm }^{-1}.
\ee
The value of $H_0$ is determined by the  dimensionless factor of
$h_{100}$ which probably lies within the range $1/2 < h_{100} < 1$.
The critical energy-density required to just close the universe is
\be \label{e:crit}
\rho_{\rm critical} = { 3 c^2 H_0^2 \over 8 \pi G} \approx 1.6 \times
10^{-8} h_{100}^2 \rm \> ergs/cm^3.
\ee
The spectrum of an {\it isotropic} stochastic gravitational
wave background is defined by a dimensionless function of frequency $f$
\be \label{e:defomegagw}
\Omega_{\rm gw}(f) \equiv {1 \over \rho_{\rm critical}} {d \rho_{\rm
gw} \over d \ln f}.
\ee
Here $d \rho_{\rm gw}$ is the energy-density in gravitational waves
contained within the frequency interval $(f,f+df)$.  Using the
definition $\Omega_{\rm gw}$ one obtains the relationship between the
spectrum $\Omega_{\rm gw}$ and $H(f)$.  For $f \ge 0$ one has
\be 
\Omega_{\rm gw}(f)  = {f \over \rho_{\rm critical}} {d \rho_{\rm
gw} \over df} = f { 8\pi G\over 3 c^2 H_0^2} {c^2 \over 32 \pi G}
32\pi^2 f^2 H(f) \int d\hat \Omega
 P(\hat \Omega) = {8 \pi^2 \over 3 H_0^2} f^3 H(f) \int d\hat \Omega
 P(\hat \Omega).
\ee
This formula shows the precise interpretation of $P(\hat \Omega)$. The stochastic
background energy density is made of contributions arriving from all directions
$\hat \Omega$ on the sky.  The actual value of $\Omega_{\rm gw}(f)$ is determined
by the average value of $P(\hat \Omega)$; the direction dependence of this function
is the same as the direction dependence of the arriving radiation intensity.

For this reason, we {\it define the multipole moments} $p_{\ell m}$ of
the stochastic background radiation by the expansion of $P(\hat
\Omega)$ in terms of spherical harmonic functions:
\be \label{e:pdef}
P(\hat \Omega) \equiv \sum_{\ell m} p_{\ell m} Y_{\ell m}(\hat \Omega)
\ee
where the sum is defined by
\be
\sum_{\ell m} \equiv \sum_{\ell=0}^\infty \sum_{m=-\ell}^\ell
= \sum_{m=-\infty}^\infty \sum_{\ell=|m|}^\infty.
\ee
In addition, without loss of generality we adopt the convention that
the monopole moment is normalized by the condition
\be
p_{00} \equiv \sqrt{4 \pi} \Rightarrow \int d\hat \Omega  P(\hat
\Omega) \equiv 4 \pi,
\ee
where we assume that the spherical harmonic functions are normalized in
the conventional way, so that the integrals of their squares over the
unit sphere gives unity.  Hence the spectrum of radiation is determined
entirely by $H(f)$ since for $f \ge 0$ one has
\be 
\label{e:defh}
\Omega_{\rm gw}(f)  = {32 \pi^3 \over 3 H_0^2} f^3 H(f).
\ee
The directionality of the arriving radiation is determined entirely by
the function $P(\hat \Omega)$.  Our fundamental assumption here is that
the pattern of the intensity of the stochastic background is {\it fixed
in a frame of reference at rest with respect to the cosmological
fluid.}  In other words, formula (\ref{e:pdef}) for $P(\hat \Omega)$ is
expressed in a set of coordinates $x.y,z$ which are fixed with respect
to the distant stars.  In those coordinates, the multipole moments
$p_{\ell m}$ are constants, independent of time.  The problem we
address in this paper is this: how do we determine, from the data
stream of a pair of interferometric detectors which are rotating with
the earth, the values of (or bounds on) the multipole moments $p_{\ell
m}$?

\section{Detection Strategy}
\label{s:howto}
To determine the multipole moments $p_{\ell m}$ the basic idea is to
correlate the outputs of two gravitational wave detectors, and to look
for variations of this correlated signal that are harmonics of the
earth's rotational frequency.  For this purpose, we need to consider
the relationship between two different time (or frequency) scales that
occur.

The first time scale is that defined by the light travel time $\Delta T$
between the two sites.  For the remainder of this section, we will
assume that the two sites are the Hanford and Livingston LIGO
detectors, so that $\Delta T=10.00\> \rm msec$.  The second time scale is
the period of the earth's rotation about its axis
relative to the cosmic frame, $T_{\rm e} =
8.6 \times 10^4\> {\rm sec} = 1 
{\rm\ sidereal\ day}$.  Because of the
enormous disparity between these two time scales, we can define a third
time scale, which we will refer to as the averaging time scale,
$\tau$.  We choose $\tau$ in the range
\be
\label{e:avertime}
\Delta T << \tau << T_{\rm e},
\ee
for example $\tau = 30\> \rm sec$.  It is then possible to examine
correlations between the two detectors as a function of time, averaged
over periods of length $\tau$.  Because $\tau$ is much shorter than
$T_{\rm e}$, the correlation between the two detectors will vary as the
earth rotates relative to the fixed cosmological frame, because of the
anisotropy in $P(\hat \Omega)$.  On the other hand, because $\tau$ is
much longer than the light travel time between the two detectors, and
because the detectors are sensitive to frequencies $f \approx 1/{\Delta
T}$, there is a significant correlated signal on time scales shorter
than $\tau$.

Denote the output of the first detector by
\be \label{e:sigdef1}
s_1(t) = h_1(t) + n_1(t),
\ee
where $h_1$ is the strain due to the stochastic background and $n_1$ is
the intrinsic noise of the first detector.  In similar fashion, the
output of the second detector is
\be \label{e:sigdef2}
s_2(t) = h_2(t) + n_2(t).
\ee
Let us use the subscript $i=1,2$ to label the detectors, so for example
$i=1$ denotes the Hanford, WA LIGO detector and $i=2$ denotes the
Livingston, LA LIGO detector.  The response $h_i$ of detector $i$ to
the gravitational radiation is given by
\be \label{e:strain}
h_i(t)=d_i^{ab}(t) h_{ab}(t,\vec x_i(t)),
\ee
where the position of detector $i$'s corner station is denoted by $\vec
x_i(t)$.  In this expression, the symmetric traceless tensors
$d_i^{ab}(t)$ are given by
\be \label{e:defd}
d_i^{ab}(t)={1 \over 2} \left( \hat X_i^a(t) \hat X_i^b(t) - \hat
Y_i^a(t) \hat Y_i^b(t) \right)
\ee
where the directions of detector $i$'s arms are defined by the unit
spatial vectors $\hat X_i^a(t)$ and $\hat Y_i^a(t)$. Note that {\it both
$d_i^{ab}(t)$ and $\vec x_i(t)$ are functions of time, because the
earth rotates with respect to the cosmological rest frame}.

Define quantities which are the Fourier transforms of the signals, evaluated
over an interval of one averaging time $\tau$ centered at time $t$:
\be \label{e:FT}
\tilde s_i(f,t) = \int_{t-\tau/2}^{t+\tau/2} dt' \> e^{-2 \pi i f t'} s_i(t') {\rm \ for\ } i=1,2.
\ee
These Fourier transforms are easily evaluated.  Substituting the plane wave expansion
(\ref{e:plane}) into the formula for the strain (\ref{e:strain}) and taking
the Fourier transform (\ref{e:FT}) we obtain
\be
\label{e:sig3} \tilde s_i(f,t) = \sum_A \int d\hat \Omega
\int_{-\infty}^\infty df' {\rm e}^{2 \pi i (f'-f) t} \delta_\tau(f-f')
h_A(f',\hat \Omega) d_i^{ab}(t) e_{ab}^A(\hat \Omega) {\rm e}^{-2 \pi i f' \hat
\Omega \cdot \vec x_i(t)/c} +{\rm noise\ term}.
\ee
In this expression, we have made use of the fact that the averaging
time $\tau$ is much less than the rotation period of the earth $\Delta
T_{\rm e}$, so that the vectors $X_i^a(t)$, $Y_i^a(t)$, and $\vec
x_i(t)$ may be treated as constants and taken outside of the time
integration in the Fourier transform (\ref{e:FT}).  We have also
defined the ``finite time" approximation to the Dirac delta function
\be 
\delta_\tau(f) \equiv \int_{-\tau/2}^{\tau/2} dt' \> e^{ -2 \pi i f t'}
= {\sin(\pi f \tau) \over \pi f},
\ee
which reduces to the Dirac delta function $\delta(f)$ in the limit
$\tau\to\infty$, but has the property that $\delta_\tau(0)=\tau$.
The final term on the right hand side of (\ref{e:sig3}) is linearly
proportional to the noise in detector $i$.

We now define the ``signal"
\be \label{e:sig2}
S(t) = \int_{-\infty}^{\infty} df  \tilde s_1^*(f,t) \tilde s_2(f,t)
\tilde Q(f)
\ee
where $\tilde Q(f)$ is an optimal filter function, to be determined.
Let us now determine the expectation value of $S(t)$ and show how it
incorporates information about the multipole moments of the stochastic
background.  To find the expected value $S(t)$ we begin by assuming
that the noise in each detector has zero mean value, and is
uncorrelated with noise and gravitational strain in the other
detector.  Under these assumptions, we find
\beau
\langle S(t) \rangle &= &
\int_{-\infty}^{\infty} df \tilde Q(f)
 \sum_A  \sum_{A'} \int_{-\infty}^{\infty} df' \int_{-\infty}^{\infty} df''
\int d\hat \Omega \int d\hat \Omega' 
{\rm e}^{-2 \pi i (f'-f) t} {\rm e}^{2 \pi i (f''-f) t} \times \\
& & \delta_\tau(f-f') \delta_\tau(f-f'') d_1^{ab}(t) d_2^{cd}(t)
 e_{ab}^A(\hat \Omega) e_{cd}^{A'}(\hat \Omega')
{\rm e}^{2 \pi i (f' \hat \Omega \cdot \vec x_1(t) - f''\hat \Omega' \cdot \vec x_2(t))/c} \times \\
& & \langle  h_A^*(f',\hat \Omega) h_{A'}(f'',\hat \Omega') \rangle.
\eeau
We now substitute in the expectation value for the product of the
amplitudes (\ref{e:quad}).  The integration over $f''$ is now trivial.
In the resulting expression, because $1/\tau$ is much smaller than the
``bandwidth" $1/\Delta T$ of the signals, one of the finite-width delta
functions $\delta_\tau$ may be replaced by a Dirac delta function. The
integration over $f'$ is then trivial.  The other finite-width delta
function is then evaluated at zero argument, giving rise to a factor of
$\tau$.  One thus obtains
\be \label{e:expectedsig}
\langle S(t) \rangle = \tau d_1^{ab}(t) d_2^{cd}(t)
\int_{-\infty}^{\infty} df \tilde Q(f) H(f) 
\int d\hat \Omega P(\hat \Omega) {\rm e}^{2 \pi i f \hat
\Omega \cdot \Delta \vec x(t)/c}
 \sum_A  e_{ab}^A(\hat \Omega) e_{cd}^{A}(\hat \Omega)
\ee
where $\Delta \vec x(t) = \vec x_1(t) - \vec x_2(t)$ is the
time-dependent separation vector between the two interferometer sites.

Not surprisingly, this previous expression can be easily simplified for
the isotropic case $P(\hat \Omega)=1$.  In this instance, the sum over
polarizations and integral over directions can be performed explicitly,
yielding ($8 \pi/5$ times) a time-independent function of frequency
known as the overlap reduction function $\gamma(f)$.  This overlap
reduction function is given by
\be
\label{e:overlap}
\gamma(f) \equiv {5 \over 8 \pi} \; d_1^{ab}  d_2^{cd} \int_{S^2} d \hat \Omega \>
e^{2 \pi i f \hat \Omega \cdot \Delta \vec x/c }
 \left(
e_{ab}^+(\hat \Omega) e_{cd}^{+}(\hat \Omega)  +
e_{ab}^\times(\hat \Omega) e_{cd}^{\times}(\hat \Omega)\right).
\ee
Notice that in (\ref{e:overlap}) the dependence of the positions and
orientations of the detectors upon time $t$ is {\it not} shown; this is
because $\gamma(f)$ depends only upon the {\it relative} positions and
orientations, which is time (or earth-position) independent.  Thus, in
the case of an isotropic stochastic background, one finds
\be
P(\hat \Omega)=1 \Rightarrow
\langle S(t) \rangle = {8 \pi \over 5} \tau \int_{-\infty}^{\infty} df \tilde Q(f) H(f) \gamma(f).
\ee
This is equation (30) of reference \cite{myreview}.  In the present paper, we are
most interested in the anisotropic case where $P(\hat \Omega)$ varies with
direction.  In this case, the time variation of the tensors
$d_i^{ab}(t)$ and $\Delta \vec x(t)$ will provide a time-dependent
variation of the signal $S(t)$.

\section{Rotation Harmonics}
\label{s:rothar}
Because the rotation of the earth is periodic with period $T_{\rm e}$
and angular frequency $\omega_{\rm e}=2 \pi/T_{\rm e}$ the expected
signal (\ref{e:expectedsig}) varies with the same period.  It can
therefore be represented by the Fourier series
\be
\label{e:fourier}
\langle S(t) \rangle = \sum_{m=-\infty}^\infty \langle S_m \rangle {\rm
e}^{i m \omega_{\rm e} t}.
\ee
Because the signal is real, the amplitudes of the different harmonics
satisfy $S_m = S^*_{-m}$.  The amplitudes are quantities which would be
determined by Fourier transforming the actual data:
\be
\label{e:harmdef}
S_m = {1 \over T} \int_0^T dt \>
  {\rm e}^{-i m \omega_{\rm  e} t} S(t).
\ee
Here $T$ represents the total observation time, which later enters into
(\ref{e:nm}) and which is assumed to be is a multiple of the earth's
rotation period $T_{\rm e}$.
The harmonic amplitudes $S_m$ are the (at least in principle)
observable quantities on which any data analysis must be based; it is
their expected values $\langle S_m \rangle $ which arise in the formula
for the expected signal.  Note that in an actual observation or
measurement, instrumental noise in the gravitational-wave detectors
would prevent $S_m$ from taking on its {\it expected} value.  In
Section~\ref{s:sigtonoise} we analyze the typical deviations of $S_m$
from $\langle S_m \rangle$ and in this way determine how accurately
$S_m$ may in fact be measured.

Because we have assumed that the $z$-axis of our (cosmic) coordinate
system points along the direction of the earth's axis, the $m$'th
rotation harmonic can only result from anisotropies whose phase varies
with angle $\phi$ as $\exp(i m \phi)$.  These are the anisotropies
associated with the $Y_{\ell m}$.  Hence
\be
\label{e:meansm}
{\langle S_m \rangle} = {8 \pi \over 5} \tau \int_{-\infty}^{\infty} df
\tilde Q(f) H(f) \sum_{\ell=|m|}^\infty p_{\ell m} \gamma_{\ell m}(f)
\ee
The functions $\gamma_{\ell m}(f)$ are generalizations of the overlap
reduction function $\gamma(f)$, which express the (frequency-dependent)
contribution of the $\ell$'th multipole moment to the $m$'th harmonic
of the signal, with respect to the Earth's rotation.  These are given
by
\be \label{e:rhocoeff}
\gamma_{\ell m}(f) = 
{5 \over 8 \pi} {1 \over 2 \pi} \int_0^{2 \pi} d\alpha \;  {\rm e}^{-i m
\alpha} d_1^{ab}(\alpha) d_2^{cd}(\alpha)
\int d\hat \Omega \; Y_{\ell m}(\hat \Omega) {\rm e}^{2 \pi i f \hat \Omega
\cdot \Delta \vec x(\alpha)/c}   
\sum_A  e_{ab}^A(\hat \Omega) e_{cd}^{A}(\hat \Omega).
\ee
In this expression, the angle of rotation of the earth about its axis
(measured from some arbitrary fiducial point) is denoted by $\alpha \in
[0,2\pi)$ so $\alpha = \omega_e t + {\rm constant} \> \pmod{2\pi}$.
The ``time-dependent" quantities $d_i^{ab}$ and $\Delta \vec x$ may
equivalently be expressed as functions of $\alpha$.

The problem at hand is now a mathematical one -- to calculate the
functions $\gamma_{\ell m}(f)$ which are generalizations of the overlap
reduction function $\gamma(f)$. For the monopole moment ($\ell=m=0$) it
is easy to see that the integrand above is independent of
earth-position $\alpha$ because the overlap reduction function
(\ref{e:overlap}) only depends upon the relative orientations of the
detectors, which is $\alpha$-independent, giving
\be
\gamma_{0 0}(f) = (4 \pi)^{-1/2} \gamma(f) .
\ee
In the next parts of this paper, we will show how to evaluate the other
$\gamma_{\ell m}$.

Our first task is to evaluate the integrals that appear in
(\ref{e:rhocoeff}).  The product $d_1^{ab}(\alpha) d_2^{cd}(\alpha)$ is a
quartic polynomial in $\sin \alpha$ and $\cos \alpha$.  One approach would
be to attempt to perform the integral over $\hat \Omega$, to obtain the
resulting function of $\alpha$, and then to evaluate the integral over
$\alpha$.  However this approach is rather cumbersome.

A more promising method is to consider the projector onto the plane
perpendicular to $\hat \Omega$, which may be calculated in terms of the vectors
defined by (\ref{e:mdef}) and (\ref{e:ndef}):
\be
Q_{ab} = \delta_{ab}-\hat \Omega_a \hat \Omega_b = {\hat
m}_a {\hat m}_b + {\hat n}_a {\hat n}_b.
\ee
A couple minutes of algebra starting with (\ref{e:polar}) and
(\ref{e:polar2}) quickly establishes the identity
\be
\sum_A  e_{ab}^A(\hat \Omega) e_{cd}^{A}(\hat \Omega) = Q_{ac} Q_{bd} +
Q_{ad} Q_{bc} - Q_{ab} Q_{cd}.
\ee
We then define the set of integrals
\be \label{e:definec}
C_{abcd}(\alpha) = 
\int d\hat \Omega \; Y_{\ell m}(\hat \Omega) {\rm e}^{2 \pi i f \hat \Omega
\cdot \Delta \vec x(\alpha)/c}  \hat \Omega_a \hat \Omega_b \hat \Omega_c \hat \Omega_d.
\ee
The desired integrals can then be expressed in terms of this quantity.
For convenience, we introduce a symbol to handle the contractions that
occur.  This is a constant tensor defined by
\be \label{e:defthe}
\Theta_{abcd}^{pqrs} =
2 \delta_{ac} \delta_{bd}  \delta^{pq} \delta^{rs} 
-4 \delta_{ac} \delta^{pq} \delta_b{}^r \delta_d{}^s 
+ \delta_a{}^p \delta_b{}^q \delta_c{}^r \delta_d{}^s.
\ee
Making use of the fact that each of the $d_i^{ab}$ is symmetric in its
tensor indices, and traceless, we may then write
\be
\label{e:messy}
\gamma_{\ell m}(f) = 
{5 \over 8 \pi} {1 \over 2 \pi} \int_0^{2 \pi} d\alpha \;  {\rm e}^{-i m
\alpha} d_1^{ab}(\alpha) d_2^{cd}(\alpha) \Theta_{abcd}^{pqrs} C_{pqrs}(\alpha).
\ee
From this definition it is easy to show that
$\gamma_{\ell,-m} = (-1)^{\ell + m} \gamma^*_{\ell,m}$.  This follows
from the parity tranformation property of the spherical harmonics
$Y_{\ell m}({-  {\hat \Omega}}) = (-1)^\ell Y_{\ell m} ({\hat
\Omega})$.  In order to now evaluate $C_{abcd}$ it is convenient to
introduce some additional coordinate systems.

\section{Coordinate Frames}
\label{s:frames}
The vectors being used in this calculation are three-dimensional
spatial vectors in flat Cartesian $\rm R^3$. Up to this point, we have
been using a coordinate system which is fixed with respect to the
cosmological fluid, and in which the spatial pattern of the
perturbations of the stochastic background is assumed to be
time-independent.  This frame of reference is the ``unprimed" frame;
vectors expressed with respect to these cosmic coordinates have
unprimed indices. We have also assumed (without any loss of generality)
that the $z$-axis of this cosmic frame points along the direction of
the earth's rotation axis.

At this point, for calculational purposes, it is convenient to consider
two additional coordinate systems.  Thus, a given vector $V$ may be
expressed in terms of its components in three different frames:
\beau
{\rm Cosmic\ Frame} & : & \quad V^a \cr
{\rm Earth\ Frame} & : & \quad V^{\bar a} \cr
{\rm Computational\ Frame} & : & \quad V^{a'}
\eeau 
The ``earth frame" is a coordinate system fixed to the earth, in which
the third ($z$-coordinate) points along the axis of the earth's
rotation, in the direction of the North pole.  Components of vectors in
this frame are denoted with ``barred" indices.  The second of these new
coordinate systems will be referred to as the ``calculational"
coordinate system.  In this frame, the components of vectors are
``primed".  This frame is fixed with respect to the earth, and has its
third ($z$-coordinate) pointing along the line between the two
gravitational-wave detectors.

The relationship between components of vectors in these three
coordinate frames may be written as matrix equations.  Each of the
matrices which appears is a special case of a rotation matrix which may
be parametrized by Euler angles.  Throughout this paper, we use the
Euler angle conventions given by equations (4.83-6) of Afkin
\cite{Arfkin} which are also the conventions used in equations (4.5)
and (4.43) of Rose \cite{rose}.  It is convenient to define a pair of
rotations about the $z$ and $y$ axes
respectively, by
\be
{\bf R}_z(\alpha) \equiv \pmatrix{ \cos \alpha & \sin \alpha & 0 \cr
                          -\sin \alpha & \cos \alpha & 0 \cr
                              0      &         0 & 1}
\qquad
{\bf R}_y(\beta) \equiv \pmatrix{ \cos \beta   & 0  & -\sin \beta   \cr
                                   0     &  1 & 0            \cr
                              \sin \beta &  0 & \cos \beta }
\ee
The most general possible rotation may be parametrized by Euler angles
and is defined by the matrix ${\bf R}(\alpha,\beta,\gamma) = {\bf R}_z(\gamma)
{\bf R}_y(\beta) {\bf R}_z(\alpha)$.  Note that the boldface symbols
here denote $3\times3$ square matrices.

The matrix which relates components of vectors in the cosmic and earth
frames is simply rotation through angle $\alpha$ about the $z$-axis:
\be \label{e:tran1}
X^{\bar a} = R^{\bar a}{}_a X^a \qquad R^{\bar a}{}_a = {\bf
R}(\alpha,0,0) = {\bf R}_z(\alpha)
\ee
where the first index on $R$ labels rows and the second index labels
columns, so that the operation appearing in the previous equation is
ordinary multiplication of a column vector on the right by a square
matrix on the left.  Note that the angle $\alpha = \omega_{\rm e} t$
varies with time.

Without loss of generality, assume that the freedom to choose the $\bar
x$- and $\bar y$-axis in the Earth frame has been used to ensure that
in this frame the separation vector $\Delta x^{\bar a}$ between the two
detector sites has no $\bar y$-component.  Using the two LIGO sites as
an example, the Earth-frame $\bar x$-axis would point out from the
center of the earth at an angle $38.6881^\circ$ East of the $0^\circ$
line of longitude (Greenwich, England).  In this frame, the coordinates
of the two detector sites and the detector arms directions are
\beau
{\rm Hanford,\ Washington:\ }
x_1^{\bar a}&=&\pmatrix{707.41 \cr -4329.11 \cr 4614.74 }^{\bar a} {\rm km,} \quad
\hat X_1^{\bar a} = \pmatrix{-0.684779 \cr 0.476172 \cr 0.55167}^{\bar a}, \quad
\hat Y_1^{\bar a} = \pmatrix{-0.720231 \cr -0.557622 \cr -0.412703}^{\bar a} \cr
{\rm Livingston,\ Louisiana:\ }
x_2^{\bar a}&=&\pmatrix{3371.80 \cr -4329.11 \cr 3240.36}^{\bar a} {\rm km,} \quad
\hat X_2^{\bar a} =  \pmatrix{-0.65377 \cr -0.708366 \cr -0.266085}^{\bar a}, \quad
\hat Y_2^{\bar a} = \pmatrix{0.540953 \cr -0.191642 \cr -0.81893}^{\bar a}.
\eeau
It is obvious that, as claimed, the separation vector between the two
sites $x_1^{\bar a} - x_2^{\bar a}$ has vanishing $\bar y$-component.  The
matrix which relates the components of the vectors in the computational
and earth frames is a rotation about the $\bar y$ axis:
\be \label{e:tran2}
X^{\bar a} = R^{\bar a}{}_{a'} X^{a'} \qquad 
R^{\bar a}{}_{a'}  = {\bf R}(0,-\beta,0)= {\bf R}_y(-\beta)
\ee
where $\beta$ is a time-independent (or $\alpha$-independent) angle,
determined by the relative orientation of the line between the two
detector sites and the earth's axis. For the two LIGO detectors, the
angle relating the Earth frame and the computational frame is
\be
\beta=-62.71383^\circ.
\ee
Within the computational frame,
the separation vector between the two sites is
\be
x_1^{a'} - x_2^{a'}  =  2997.98 \>{\rm km} \pmatrix{0\cr 0\cr 1}
\ee
and the unit-length vectors defining the arm directions are
\be \label{e:armdir}
\hat X_1^{a'} = \pmatrix{0.176358\cr 0.476171\cr 0.861486}^{a'}\qquad 
\hat Y_1^{a'} = \pmatrix{-0.69696\cr -0.557623\cr 0.450893}^{a'}\qquad 
\hat X_2^{a'} = \pmatrix{-0.536188\cr -0.708366\cr 0.459042}^{a'}\qquad 
\hat Y_2^{a'} = \pmatrix{-0.479814\cr -0.191642\cr -0.856185}^{a'}_.
\ee
These quantities will become useful later.

\section{Computation of $C$}
\label{s:ccomp}
Our goal now is to calculate $C_{abcd}$ as defined in
(\ref{e:definec}).  To do this, we will express the spherical harmonic
functions $Y_{\ell m}(\Omega) = Y_{\ell m}(\theta,\phi)$ in terms of
the ``primed" coordinates in the computational frame.
Combining the transformations (\ref{e:tran1}) and (\ref{e:tran2})
we obtain the relationship between vectors in the cosmic and computational
frames:
\be
V^{a'} = {\bf R}_y(\beta) {\bf R}_z(\alpha) V^{a} = {\bf R}(\alpha,\beta,0) V^{a}.
\ee
This transformation through Euler angles $\alpha,\beta,0$ induces a
simple change in the spherical harmonics.  For a given value of $\ell$
the spherical harmonic functions in one frame are simply a sum of all
the spherical harmonics with the same value of $\ell$ in the other
frame.  The relation between these two sets of functions is given by
the rotation matrices $D^\ell_{mn}$, which are closely related to
Clebsch-Gordon coefficients:
\be
Y_{\ell m}(\theta',\phi') = \sum_{k=-\ell}^\ell D^\ell_{km}(\alpha,\beta,0) Y_{\ell k}(\theta,\phi).
\ee
(See equation (4.260) of Arfkin \cite{Arfkin}.)
The inverse transformation is obtained by reversing the lower two indices on the
rotation matrix and complex-conjugating:
\be
Y_{\ell m}(\theta,\phi) = \sum_{k=-\ell}^\ell \left( D^\ell_{mk}(\alpha,\beta,0)
\right)^* Y_{\ell k}(\theta',\phi').
\ee
The rotation matrices are conveniently expressed by equation (4.12) of
Rose \cite{rose}:
\be
D^\ell_{mk}(\alpha,\beta,\gamma)={\rm e}^{-i m \alpha}
d^\ell_{mk}(\beta) {\rm e}^{-i k \gamma}.
\ee
Explicit formulae for the $d^\ell_{mk}$ may be given either in the form
of a sum, or in ``summed" form.  The latter expression, given by
equation (4.14) of Rose \cite{rose} is the most useful one for us. 
For $m \ge k$
\be
d^\ell_{mk} =
\left[{ (\ell - k)! (\ell + m)! \over (\ell + k)! (\ell - m)! } \right]^{1/2}
{ \left( \cos {\beta \over 2} \right)^{2\ell + k - m} 
  \left( -\sin {\beta \over 2} \right)^{m-k} \over (m-k)! } \>
{}_2 F_1(m-\ell,-k-\ell;m-k+1;-\tan^2 {\beta \over 2} ). \label{dlmk}
\ee
Notice that because the $m \le \ell$ the first argument of the Gauss
hypergeometric function ${}_2F_1$ is a non-positive integer
  the hypergeometric series ${}_2F_1$ terminates after a
finite number of terms.  In fact it is possible to rewrite
equation Eq. (\ref{dlmk}) in terms of Jacobi polynomials
$P_n^{(\alpha,beta)}$, for $m \ge k$
\be
d^\ell_{mk} = (-1)^{l-m}
\left[{ (\ell - m)! (\ell + m)! \over (\ell - k)! (\ell +k)! } \right]^{1/2}
{ \left( \cos {\beta \over 2} \right)^{m+k} 
  \left( -\sin {\beta \over 2} \right)^{m-k}} \>
  P_{\ell-m}^{(m+k,m-k)}(-\cos\beta).
\ee
In the event that $m < k$ the $d^\ell_{mk}$
may be obtained from the unitarity property, equation (4.15) of Rose
\cite{rose}
\be
d^\ell_{mk}(\beta) = d^\ell_{km}(-\beta)
= (-1)^{m-k} d^\ell_{km}(\beta).
\ee
Note also that the $d^\ell_{mk}$ are real, so that we can drop the
complex conjugation that would otherwise have appeared.

The integral over the two-sphere which appears in (\ref{e:definec}) can
also be expressed as an integral over all directions in the
computational (primed) frame. In other words,
$\int d\hat \Omega = \int d\hat \Omega'$.
So our integral may be expressed as
\be
C_{abcd}(\alpha) =   \sum_{k=-\ell}^\ell d^\ell_{mk}(\beta) {\rm e}^{i
m \alpha} N_\ell^k \int_0^\pi \sin \theta' d\theta' {\rm e}^{2 \pi i f
\Delta T \cos \theta'} {\rm P}^k_\ell( \cos \theta') \int_0^{2 \pi}
d\phi' {\rm e}^{i k \phi'} \hat \Omega_a \hat \Omega_b \hat \Omega_c
\hat \Omega_d,
\ee
where we have expressed the spherical harmonic functions in terms of
associated Legendre functions $\rm P_\ell^k$, and $\Delta T =|x_1 -
x_2|/c$ denotes the light travel time between the two detector sites
(10.00 msec for the two LIGO detectors).  The normalization constants
$N^k_\ell$ which relate the spherical harmonics and the Legendre
polynomials are
\be
N^k_\ell = \sqrt{ {2 l +1 \over 4 \pi} {(l-k)! \over (l+k)!}}.
\ee
We will eventually be contracting the four indices of $C$ with each
other and with the indices of other tensors.  Of course such
contractions yield the same result in any coordinate frame, and it is
easier to calculate $C$ in the calculational (primed) frame.  Hence we
may write $C_{abcd}(\alpha) = R_a{}^{{ a'}}
R_b{}^{{b'}}R_c{}^{{c'}}R_d{}^{{d'}} C_{{a'} {b'} {c'} {d'}}$ where
$R_a{}^{{a'}} = {\bf R}_z(-\alpha) {\bf R}_y(-\beta)$, and
\be \label{e:primeform}
C_{a'b'c'd'} =   {\rm e}^{i m \alpha} \sum_{k=-\ell}^\ell
d^\ell_{mk}(\beta) N_\ell^k
\int_0^\pi \sin \theta' d\theta' {\rm e}^{2 \pi i f \Delta T \cos \theta'}
{\rm P}^k_\ell( \cos \theta') \int_0^{2 \pi} d\phi' {\rm e}^{i k \phi'}
\hat \Omega_{a'} \hat \Omega_{b'} \hat \Omega_{c'} \hat
\Omega_{d'}.
\ee
The vector $\hat \Omega^{a'}$ is
\be \label{e:omegaprime}
\hat \Omega^{a'} =
\pmatrix{ \cos \phi' \sin \theta' \cr \sin \phi' \sin \theta' \cr
\cos \theta'}^{a'}.
\ee
It is clear that the integral over $\phi'$ in (\ref{e:primeform})
vanishes unless $k=-4,-3,\cdots,3,4$.   Thus even for large $\ell$ the
range of summation over $k$ only includes these values.  There is a
sense in which this reflects the fact that our signal is a product of
the outputs of a pair of detectors, each of which has a quadrupole
antenna pattern.  It is also noteworthy that the remaining integral,
over the variable $\theta'$, can also be done explicitly for any
distinct values of $\ell$ and $k$.

\section{The remaining integrations}
\label{s:integ}
We are now in a position to evaluate the remaining integrals.  We begin
with the integral over $\alpha$.  We can re-write $\gamma_{\ell m}$ from
(\ref{e:messy}) as
\beau
\gamma_{\ell m}(f) &= &
{5 \over 8 \pi} {1 \over 2 \pi} \int_0^{2 \pi} d\alpha \;  {\rm e}^{-i m
\alpha} {\rm e}^{i m \alpha} d_1^{a'b'}(\beta) d_2^{c'd'}(\beta) \times \\
&& \Theta_{a'b'c'd'}^{p'q'r's'}
\sum_{k=-\ell}^\ell
d^\ell_{mk}(\beta) N_\ell^k
\int_0^{\pi}  \sin \theta' d\theta' \; {\rm e}^{2 \pi i f \Delta T \cos \theta'}
{\rm P}^k_\ell( \cos \theta') \int_0^{2 \pi} d\phi' {\rm e}^{i k \phi'}
\hat \Omega_{p'} \hat \Omega_{q'} \hat \Omega_{r'} \hat \Omega_{s'}.
\eeau
In this integral, we have explicitly indicated all of the dependence on
$\alpha$.  Notice that while $d_i^{ab}$ is a function of $\alpha$, in
computational coordinates $d_i^{a'b'}(\beta)$ is independent of
$\alpha$ and depends only upon $\beta$.  Likewise, the tensor
$\Theta_{a'b'c'd'}^{p'q'r's'}$ defined by (\ref{e:defthe}) has constant
components in the computational frame.  Hence the integral over
$\alpha$ give a factor $2 \pi$:
\bea \label{e:simpler}
\gamma_{\ell m}(f) &= &
{5 \over 8 \pi} \; d_1^{a'b'}(\beta) d_2^{c'd'}(\beta)  \times \cr
& & \Theta_{a'b'c'd'}^{p'q'r's'}
\sum_{k=-\ell}^\ell
d^\ell_{mk}(\beta) N_\ell^k 
\int_0^{\pi}  \sin \theta' d\theta' \; {\rm e}^{2 \pi i f \Delta T \cos \theta'}
{\rm P}^k_\ell( \cos \theta') \int_0^{2 \pi} d\phi' {\rm e}^{i k \phi'}
\hat \Omega_{p'} \hat \Omega_{q'} \hat \Omega_{r'} \hat \Omega_{s'}.
\eea
The form of this integral is interesting.  The integral over $\phi'$
will vanish unless $k=-4,\cdots,4$, in which case it yields a product
of at most four factors of $\sin \theta'$ and $\cos \theta'$.
Introducing a new variable $u=\cos \theta'$ the two integrals appearing
in (\ref{e:simpler}) may be expressed as linear combinations of
the integrals
\be
\int_{-1}^1  du \; {\rm e}^{2 \pi i f \Delta T u}
{\rm P}^k_\ell( u) u^N (1-u^2)^{|k|/2}
\label{u_integral}
\ee
where $N$ is a  non-negative integer bounded by $N+|k| \le 4$. 
Such integrals can be expressed in closed-form as we show in the Appendix.

\begin{figure}
\begin{center}
\epsfig{file=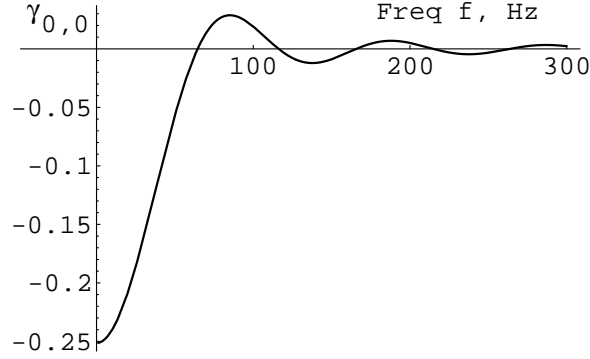,width=3.1in,bbllx=2.8cm,bblly=11.4cm,
bburx=10.5cm,bbury=16.3cm}
\end{center}
\caption{
\label{f:fig0}
The (real) function $\gamma_{0,0}(f)$ is shown for the LIGO pair of
detectors.}
\end{figure}

\begin{figure}
\begin{center}
\epsfig{file=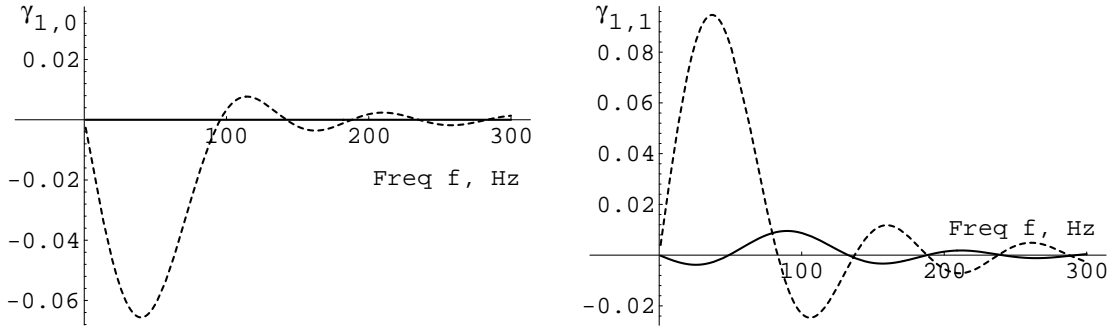,width=6.0in,bbllx=2.8cm,bblly=11.4cm,
bburx=19.2cm,bbury=16.5cm}
\end{center}
\caption{
\label{f:fig1}
The functions $\gamma_{1,m}(f)$ are shown for the LIGO pair of
detectors.  The real parts are shown as the solid curves, and the imaginary
parts as the dotted curves.}
\end{figure}

\begin{figure}
\begin{center}
\epsfig{file=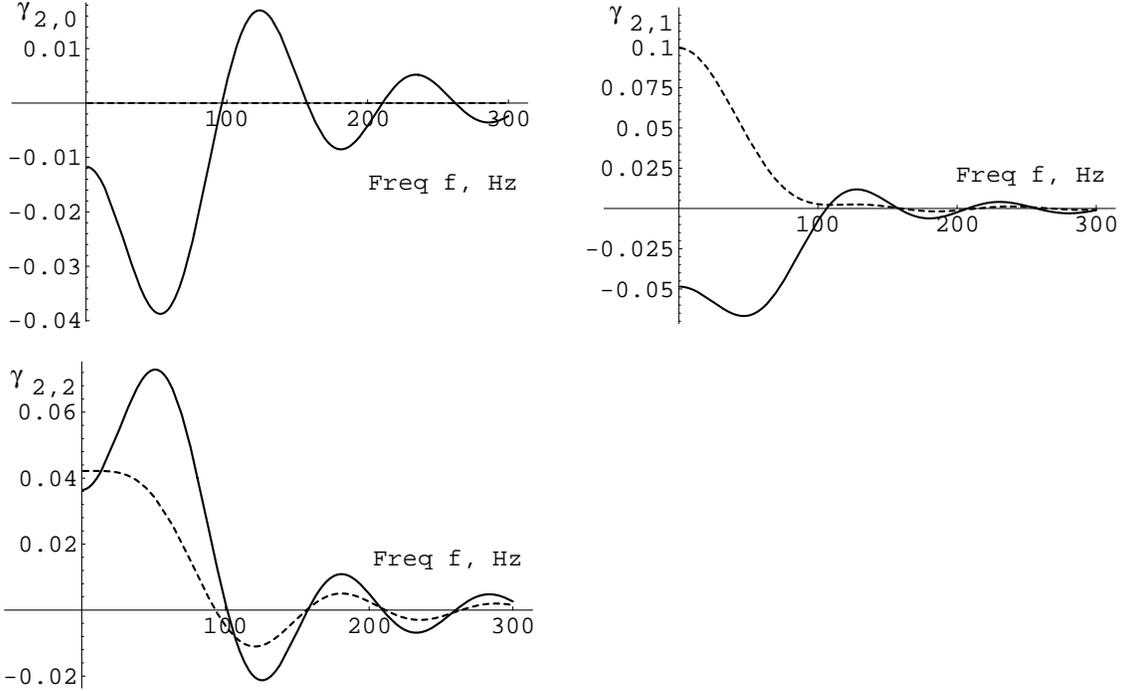,width=6.0in,bbllx=2.8cm,bblly=8.9cm,
bburx=19cm,bbury=19cm}
\end{center}
\caption{
\label{f:fig2}
The functions $\gamma_{2,m}(f)$ are shown for the LIGO pair of
detectors.  The real parts are shown as the solid curves, and the imaginary
parts as the dotted curves.}
\end{figure}

\begin{figure}
\begin{center}
\epsfig{file=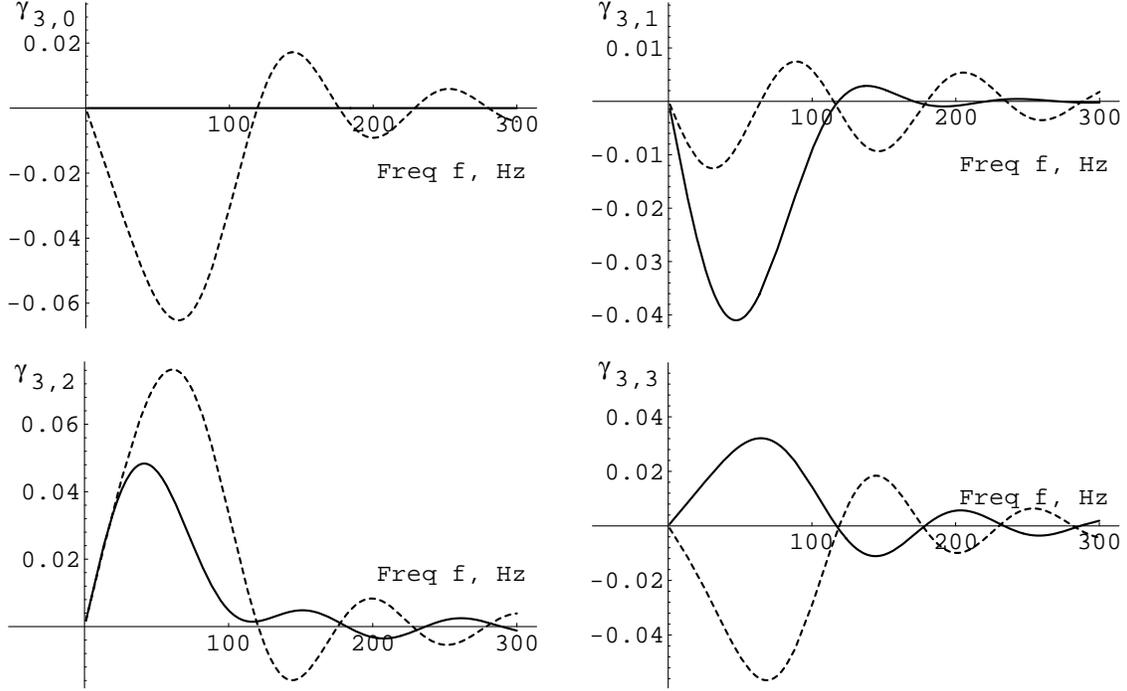,width=6.0in,bbllx=2.8cm,bblly=8.9cm,
bburx=19cm,bbury=19cm}
\end{center}
\caption{
\label{f:fig3}
The functions $\gamma_{3,m}(f)$ are shown for the LIGO pair of
detectors.  The real parts are shown as the solid curves, and the imaginary
parts as the dotted curves.}
\end{figure}

From this point on, we consider only the case of the two LIGO
detectors.  In this case the simplest way to proceed is to compute
\be
s_k(\theta')=d_1^{a'b'}(\beta) d_2^{c'd'}(\beta) \Theta_{a'b'c'd'}^{p'q'r's'}
\int_0^{2 \pi} d\phi' {\rm e}^{i k \phi'} \hat
\Omega_{p'} \hat \Omega_{q'} \hat \Omega_{r'} \hat \Omega_{s'}.
\ee
To evaluate this, we use definition (\ref{e:defd}) of the $d_i^{ab}$, the contraction
operator (\ref{e:defthe}) and the arm directions (\ref{e:armdir}).  
Substituting in the vector $\hat \Omega^{a'}$ given by (\ref{e:omegaprime}) gives
elementary integrals over $\phi'$.  The results are easily written in terms of
the variable $u=\cos \theta'$:
\beau
s_0 (u) & = & -3.01308 + 1.75421\,{u^2} + 0.945109\,{u^4} \cr
s_1 (u) & = &   \left( 0.5451 + 0.543353\,i 
     -\left( 1.6954 + 1.73284\,i \right) \,{u^2} \right) \,u \,(1 - {u^2})^{1/2} \cr
s_2 (u) & = &  \left( -0.0428705 + 1.41125\,i + \left( 0.0119604 -
     0.670812\,i \right) \,{u^2} \right) (1-u^2) \cr
s_3 (u) & = &  (-0.245744 + 0.227815i) \,u \,(1-u^2)^{3/2} \cr
s_4 (u) & = &  (0.0492709 + 0.00345516 i) \, (1-u^2)^2  \cr
s_{-k} (u) & = & s_{k}^* (u)\cr
s_k (u) & = & 0 \quad {\rm for\ } |k|>4
\eeau
We then have
\be
\gamma_{\ell m}(f) = {5 \over 8 \pi} \; \sum_{k=-\ell}^\ell
d^\ell_{mk}(\beta) N_\ell^k \int_{-1}^1  du \; {\rm
e}^{2 \pi i f \Delta T u} {\rm P}^k_\ell(u) s_k(u).
\ee
We now evaluate these functions for the first few multipoles.  For this
purpose we introduce a dimensionless frequency variable $x=2 \pi f \Delta T$.  Because $\gamma_{\ell m} = (-1)^{\ell+m}
\gamma^*_{\ell,-m}$ we give these functions only for $m=0,\cdots,\ell$.
They may be conveniently written in terms of spherical Bessel functions
$j_n$.  For $\ell=0$ one has
\beu
\gamma _{0,0}(x) = -0.0352174 j_{0}(x) - 0.818115 {{j_{1}(x)}\over x} + 
    0.848647 {{j_{2}(x)}\over {{x^2}}}.
\eeu
A graph of this function is shown in Figure~\ref{f:fig0}.
For $\ell=1$ one has
\beau
\gamma _{1,0}(x) & = & -0.0279637 i j_{1}(x) - 0.252119 i {{j_{2}(x)}\over x} - 
               1.66955 i {{j_{3}(x)}\over {{x^2}}}\cr
\gamma _{1,1}(x) & = & 0.0383329 i j_{1}(x) - 
 (0.327033 - 1.03547 i) {{j_{2}(x)}\over x} + 
 (1.90568 - 1.77847 i) {{j_{3}(x)}\over {{x^2}}}.
\eeau
A graph of these functions is shown in Figure~\ref{f:fig1}.
For $\ell=2$ one has
\beau
\gamma _{2,0}(x) & = & 0.0145494 j_{0}(x) + 1.00009 {{j_{1}(x)}\over x} - 
    9.39901 {{j_{2}(x)}\over {{x^2}}} + 28.0344 {{j_{3}(x)}\over {{x^3}}}\cr
\gamma _{2,1}(x) & = & 0.0392947 j_{0}(x) + 
    (0.385015 + 0.335238 i) {{j_{1}(x)}\over x} - 
    (2.38288 + 2.48534 i) {{j_{2}(x)}\over {{x^2}}}  \cr
 & & - (6.01443 - 16.1191 i) {{j_{3}(x)}\over {{x^3}}}\cr
\gamma _{2,2}(x) & = & 
-0.0380887 j_{0}(x) - (1.05867 + 0.649899 i) {{j_{1}(x)}\over x} + 
 (9.05483 + 6.80403 i) {{j_{2}(x)}\over {{x^2}}} \cr
& & - (18.5056 + 20.4538 i) {{j_{3}(x)}\over {{x^3}}}.
\eeau
A graph of these functions is shown in Figure~\ref{f:fig2}.
For $\ell=3$ one has
\beau
\gamma _{3,0}(x) & = & 0.0416301 i j_{1}(x) + 0.805209 i {{j_{2}(x)}\over x} - 
    11.2779 i {{j_{3}(x)}\over {{x^2}}} + 18.399 i {{j_{4}(x)}\over {{x^3}}}
   \cr
\gamma _{3,1}(x) & = & 0.00182185 i j_{1}(x) - 
    (0.0155429 + 1.03107 i) {{j_{2}(x)}\over x} - 
    (1.38339 - 11.1549 i) {{j_{3}(x)}\over {{x^2}}} \cr & & - 
    (7.72558 + 45.6531 i) {{j_{4}(x)}\over {{x^3}}}\cr
\gamma _{3,2}(x) & = & -0.0461979 i j_{1}(x) + 
    (0.788263 - 0.534988 i) {{j_{2}(x)}\over x} - 
    (7.13839 - 8.50643 i) {{j_{3}(x)}\over {{x^2}}}\cr  & & + 
    (42.9458 - 1.80521 i) {{j_{4}(x)}\over {{x^3}}}\cr
\gamma _{3,3}(x) & = & 0.0365627 i j_{1}(x) - 
    (0.935792 - 1.08142 i) {{j_{2}(x)}\over x} + 
    (11.3339 - 13.8287 i) {{j_{3}(x)}\over {{x^2}}} \cr & & - 
    (33.6244 - 32.4339 i) {{j_{4}(x)}\over {{x^3}}}.
\eeau
A graph of these functions is shown in Figure~\ref{f:fig3}.
For $\ell=4$ one has
\beau
\gamma _{4,0}(x) & = & 0.0232299 j_{0}(x) - 0.899008 {{j_{1}(x)}\over x} + 
    9.78424 {{j_{2}(x)}\over {{x^2}}} - 88.696 {{j_{3}(x)}\over {{x^3}}} + 
    438.555 {{j_{4}(x)}\over {{x^4}}}\cr
\gamma _{4,1}(x) & = &  
-0.0367891 j_{0}(x) - 
    (1.10932 + 0.313862 i) {{j_{1}(x)}\over x} + 
    (27.3542 + 5.98172 i) {{j_{2}(x)}\over {{x^2}}} \cr & & - 
    (204.205 + 48.3679 i) {{j_{3}(x)}\over {{x^3}}} + 
    (496.286 + 149.558 i) {{j_{4}(x)}\over {{x^4}}}\cr
\gamma _{4,2}(x) & = &  -0.0155407 j_{0}(x) + 
    (1.05639 - 0.265167 i) {{j_{1}(x)}\over x} - 
    (14.276 - 0.635514 i) {{j_{2}(x)}\over {{x^2}}} \cr & & + 
    (127.588 - 28.9739 i) {{j_{3}(x)}\over {{x^3}}} - 
    (551.647 - 310.656 i) {{j_{4}(x)}\over {{x^4}}}\cr
\gamma _{4,3}(x) & = &  0.0502848 j_{0}(x) + 
    (0.521081 + 1.287 i) {{j_{1}(x)}\over x} - 
    (20.3464 + 20.2033 i) {{j_{2}(x)}\over {{x^2}}} \cr & & + 
    (135.553 + 177.464 i) {{j_{3}(x)}\over {{x^3}}} - 
    (156.429 + 719.091 i) {{j_{4}(x)}\over {{x^4}}}\cr
\gamma _{4,4}(x) & = &  -0.0344655 j_{0}(x) - 
    (0.977416 + 1.17615 i) {{j_{1}(x)}\over x} + 
    (24.322 + 22.0573 i) {{j_{2}(x)}\over {{x^2}}} \cr & & - 
    (183.738 + 163.238 i) {{j_{3}(x)}\over {{x^3}}} + 
    (458.028 + 438.49 i) {{j_{4}(x)}\over {{x^4}}}\cr
\eeau
Note that the $\gamma_{\ell m}$ with $\ell$ odd vanish as $f \to 0$, in
contrast with the functions with $\ell$ even, which approach constant
values at zero frequency.

\section{Observability and Signal-To-Noise Ratios}
\label{s:sigtonoise}

Up to this point, we have shown how anisotropies in the
gravitational-wave stochastic background give rise to periodic
variations in the ``signal" obtained by correlating a pair of
detectors.  These periodic variations are described by the Fourier
series (\ref{e:fourier}), with coefficients $S_m$.  In this section, we
address the question:  how precisely can the values of these
coefficients be determined, in the presence of noise in the two
detectors?  We answer this question by calculating the signal-to-noise ($S/N$)
ratios that would arise in measurements of the $S_m$, which also
permits us to determine the best choice of the optimal filter functions
$\tilde Q(f)$.

In carrying out this analysis, we follow the technique used in Section
3.2 of reference \cite{myreview}.  The following presentation will be
somewhat crypic as we will assume that the reader is familiar with that
material.

\begin{figure}
\begin{center}
\epsfig{file=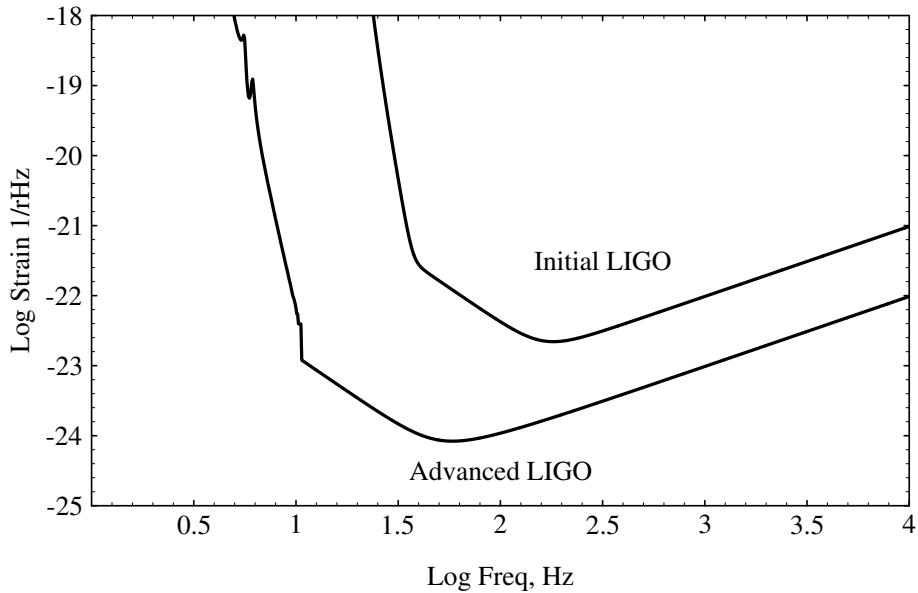, width=12cm, bbllx=74pt, bblly=248pt,
bburx=537pt, bbury=565pt}
\end{center}
\caption{
\label{f:noise}
\noindent
The predicted noise power spectra of the initial and advanced LIGO
detectors.   The horizontal axis is $\rm log_{10}$ of frequency $f$, in
Hz.  The vertical axis shows $\rm log_{10}  (P(f)/sec)^{1/2}$, or
strain per root Hz.  These noise power spectra are the published design
goals.  The bumps appearing in the low-frequency part of the advanced
LIGO noise curve are obtained by folding measured seismic noise data
with the predicted transfer function of the seismic isolation (stack)
system.}
\end{figure}

The noise in the detectors is characterized by a cross-correlation
function
\be
\label{e:nps}
\langle \tilde n_i^*(f) \tilde n_j(f') \rangle = {1 \over 2}
\delta_{ij} \delta(f-f') P_i(|f|).
\ee
Here $i=1,2$ labels the two detector sites and $P_i(f)$ is the
(one-sided, real) noise power spectrum of the $i$'th detector.  For the
initial and advanced LIGO detectors, these power spectra are shown in
Fig.~\ref{f:noise}.  For our calculations of $S/N$ we will need to know
the noise properties of the detectors averaged over our ``windowing
time" $\tau$.  To obtain these, we first characterize the detector
noise in the time domain, by Fourier transforming (\ref{e:nps}).  This
gives
\bea
\label{e:tps}
\langle n_i(t) n_j(t') \rangle &=& 
\int_{-\infty}^\infty df \> {\rm e}^{-2 \pi i f t}
\int_{-\infty}^\infty df' \> {\rm e}^{2 \pi i f' t'}
\langle \tilde n_i^*(f) \tilde n_j(f') \rangle \cr
&=&  {1 \over 2} \delta_{ij}\int_{-\infty}^\infty df \>{\rm e}^{-2 \pi
i f (t-t')} P_i(|f|).
\eea
In words, this says that the Fourier transform of the noise auto-correlation
function is the noise power spectrum.

We can now find the ``windowed" version of these formulae, using the
definition (\ref{e:FT}) of the windowed transform.  Taking the windowed
transforms of (\ref{e:tps}) yields
\bea
\langle \tilde n_i^*(f,t) \tilde n_j(f',t') \rangle &=&
\int_{t-\tau/2}^{t+\tau/2} dt'' \> {\rm e}^{2 \pi i f t''}
\int_{t'-\tau/2}^{t'+\tau/2} dt''' \> {\rm e}^{-2 \pi i f' t'''}
\langle n_i(t'') n_j(t''') \rangle \cr & = & {1 \over 2} \delta_{ij}
{\rm e}^{2 \pi i (f t - f' t')} \int_{-\infty}^\infty dp \>
{\rm e}^{-2 \pi i p (t-t')} \delta_\tau(f-p) \delta_\tau(f'-p) P_i(|p|).
\label{e:nwindow}
\eea
Note that in the long averaging time limit $\tau \to \infty$ this
reproduces equation (\ref{e:nps}).

In virtually any realistic scenario, the intrinsic instrumental detector
noise is expected to be much larger than the strain arising from the
stochastic background of gravity waves.  For this reason, if we define
the ``noise" in a given measurement of $S_m$ by $N_m \equiv S_m -
\langle S_m \rangle$ then to good approximation a formula for $N_m$
may be obtained by replacing the total detector output
$s_i(t)$ which appears in (\ref{e:sigdef1}) by $n_i(t)$.  Using
the definition of the ``signal" (\ref{e:sig2}) and the definition of
the $m$'th harmonic (\ref{e:harmdef}) one obtains
\be
\label{e:nm}
N_m = {1 \over T} \int_0^{T} dt \> {\rm e}^{-i m \omega_{\rm  e} t} 
\int_{-\infty}^{\infty} df  \tilde n_1^*(f,t) \tilde n_2(f,t)
\tilde Q(f).
\ee
(Note that in this formula, we assume that the total observation time
$T$ is large compared with the earth's rotation time $T_{\rm e}$ and
that during the observation time the earth has made an integral number
of rotations, so that $T/T_{\rm e}$ is a large integer.  We will
typically take $T$ to be 1 sidereal year.)  The approximation that we
make here is obviously consistent with $\langle N_m \rangle = 0$ since
the noise in the two detectors is assumed to be uncorrelated, so that
$\langle \tilde n_1^*(f,t) \tilde n_2(f',t') \rangle = 0$.

The ``noise" arising in the measurement of $S_m$ may be
characterized by $\langle |N_m|^2 \rangle = \langle |S_m|^2 \rangle -
| \langle S_m \rangle |^2$.  We now calculate $\langle N_m^* N_{m'}
\rangle$.  Substituting equation (\ref{e:nwindow}) into equation
(\ref{e:nm}) and ``squaring" yields
\bea
\langle N_m^* N_{m'} \rangle &=& 
{1 \over T^2} \int_0^T dt\> \int_0^T dt' \>
{\rm e}^{i \omega_{\rm e} (mt - m't') }
\int_{-\infty}^{\infty} df \> \tilde Q^*(f)
\int_{-\infty}^{\infty} df' \> \tilde Q(f')
\langle \tilde n_1^*(f,t) \tilde n_1(f',t') \rangle
\langle \tilde n_2^*(f,t) \tilde n_2(f',t') \rangle
\cr
&=&
{1 \over 4 T^2} \int_0^T dt \> \int_0^T dt'
\int_{-\infty}^{\infty} df \>
\int_{-\infty}^{\infty} df' \> 
\int_{-\infty}^{\infty} dp \>
\int_{-\infty}^{\infty} dp' \> \tilde Q^*(f) \tilde Q(f') 
{\rm e}^{ i \omega_{\rm e} (mt - m't')+ 2 \pi i (p-p')(t-t') }\cr
& &  \qquad \times  \> \delta_\tau(f-p) \delta_\tau(f'-p) \delta_\tau(f-p') \delta_\tau(f'-p') 
P_1(|p|) P_2(|p'|).
\eea
We can simplify this expression to the point where it is useful, however this
is somewhat tricky -- there is only one order in which the integrals above
can be simply evaluated to yield useful approximations.  We first do the integrals
with respect to $t$ and $t'$ exactly.  This gives
\bea
\langle N_m^* N_{m'} \rangle &=&
{1 \over 4 T^2} \int_{-\infty}^{\infty} df \>  \int_{-\infty}^{\infty} df' \> \int_{-\infty}^{\infty} dp \> \int_{-\infty}^{\infty} dp' \>
\tilde Q^*(f) \tilde Q(f') P_1(|p|) P_2(|p'|) 
{\rm e}^{-i \omega_{\rm e} T (m-m')/2} \cr
& & \qquad \times \>
\delta_T(p-p'+m\omega_{\rm e}/2 \pi) 
\delta_T(p-p'+m'\omega_{\rm e}/2 \pi) 
\delta_\tau(f-p) \delta_\tau(f'-p) \delta_\tau(f-p') \delta_\tau(f'-p') .
\eea
Now we note that the effective support of $\delta_T(f)$ extends over a
very narrow range of frequencies (typically, $|f| < 10^{-7}$ Hz) 
compared with the effective support of $\delta_\tau(f)$ 
(whose support is typically $|f| < 10^{-2}$ Hz).
In addition, none of the remaining integrand varies over such a narrow
frequency range.  So we are justified in replacing $\delta_T(p-p'+m\omega_{\rm e}/2 \pi) 
$ by the ordinary Dirac delta function $\delta(p-p'+m\omega_{\rm e}/2 \pi)$.
This gives
\bea
\langle N_m^* N_{m'} \rangle &=&
{1 \over 4 T^2} \int_{-\infty}^{\infty} df \>  \int_{-\infty}^{\infty} df' \> 
\tilde Q^*(f) \tilde Q(f') \int_{-\infty}^{\infty} dp' \>
\delta_\tau(f-p') \delta_\tau(f'-p') {\rm e}^{-i \omega_{\rm e} T (m-m')/2}
 \cr
& & \qquad \times \>
\delta_T \left( {\omega_{\rm e}(m-m') \over 2 \pi} \right) 
\delta_\tau \left( f-p'+{m\omega_{\rm e} \over 2 \pi} \right) \delta_\tau
\left( f'-p'+{m\omega_{\rm e} \over 2 \pi} \right) P_1 \left( \left| p'-{m\omega_{\rm e} \over 2 \pi} \right|
\right) P_2(|p'|).
\eea
If we now additionally assume that the observation time $T$ is much greater than the
period of a single rotation $T_{\rm e} = {2 \pi \over \omega_{\rm e}}$ then
to good approximation $\delta_T \left( {\omega_{\rm e}(m-m') \over 2 \pi} \right)
\approx T \delta_{mm'}$.  Thus
\bea
\langle N_m^* N_{m'} \rangle &=&
{1 \over 4 T} \delta_{mm'} \int_{-\infty}^{\infty} df \>  \int_{-\infty}^{\infty} df' \> 
\tilde Q^*(f) \tilde Q(f') \int_{-\infty}^{\infty} dp \>
\delta_\tau(f-p) \delta_\tau(f'-p) 
 \cr
& & \qquad \times \>
\delta_\tau \left( f-p+{m\omega_{\rm e} \over 2 \pi} \right) \delta_\tau
\left( f'-p+{m\omega_{\rm e} \over 2 \pi} \right) P_1 \left( \left| p-{m\omega_{\rm e} \over 2 \pi} \right|
\right) P_2(|p|).
\eea
Next we note that the width of the $\delta_\tau(f)$ in frequency space
is quite large compared to $m \omega_{\rm e} $ provided that we
restrict $m$ to be fairly small: $|m| < T_{\rm e}/\tau$.  We also
assume that the noise power spectrum $P_1(f)$ does not vary
significantly over frequency scales of $\omega_{\rm e} \approx 10^{-5}$
Hz.  In this case we can neglect the shifting of the arguments by $m
\omega_{\rm e} $ above, obtaining
\bea
\langle N_m^* N_{m'} \rangle &=&
{1 \over 4 }{ \tau \over T} \delta_{mm'} \int_{-\infty}^{\infty} df \>
\int_{-\infty}^{\infty} df' \>
\tilde Q^*(f) \tilde Q(f') \delta^2_\tau(f-f')  P_1 (| f |)  P_2(f|) \cr
&=& \delta_{mm'} {\tau^2 \over 4 T} 
\int_{-\infty}^{\infty} df \> |\tilde Q(f)|^2 P_1 (| f |)  P_2(| f |).
\eea
Setting $m=m'$ we finally obtain an expression for the expected
``squared noise" in a measurement of $S_m$:
\be
\label{e:meannm}
\langle  |N_m|^2 \rangle = {\tau^2 \over 4 T} 
\int_{-\infty}^{\infty} df \> |\tilde Q(f)|^2 P_1 (| f |)  P_2(| f |).
\ee
We can make use of this expression to find the optimal filter $\tilde Q(f)$.

The (squared) Signal-to-Noise ratio $S/N$ for the $m$'th harmonic is
now defined via the ratio of expected signal (magnitude squared)
divided by expected (squared) noise.  Making use of (\ref{e:meansm})
for the former quantity, and (\ref{e:meannm}) for the latter yields
\be
\left( {S \over N} \right)_m^2 \equiv 
{ | \langle S_m \rangle|^2 \over \langle |N_m|^2 \rangle}
= {  4 T \left( {8 \pi \over 5} \right)^2 
\left| \tau \int_{-\infty}^{\infty} df
\tilde Q(f)
 H(f) \begin{displaystyle}
 \sum_{\ell=|m|}^\infty \end{displaystyle} p_{\ell m} \gamma_{\ell m}(f)  
\right|^2 \over \tau^2 \int_{-\infty}^{\infty} df \> |\tilde Q(f)|^2
P_1 (| f |)  P_2(| f |) }.
\ee
Notice that the averaging time $\tau$ (which was earlier chosen in a
rather arbitrary manner) drops out of this expression.  Provided that
the assumptions  about $\tau$ (\ref{e:avertime}) used in deriving this
equation are satisfied, the actual value is irrelevant.

In order to find the optimal filter function, it is useful to introduce
an inner product.  For any complex functions of frequency $A(f)$ and
$B(f)$, this defines a complex number which is denoted by $(A,B)$.  The
definition is
\be
(A,B) \equiv \int_{-\infty}^\infty df \> A^*(f) B(f) 
P_1 (| f |)  P_2(| f |).
\ee
This inner product is positive definite because $(A,A)$ is real and non-negative,
vanishing only if $A$ is zero.  In terms of this inner product,
the Signal-to-Noise ratio may be written as
\be
\left( {S \over N} \right)_m^2 = 4 T \left( {8 \pi \over 5} \right)^2 
{  \left| \left( \tilde Q, {
H(f)  \over 
P_1 (| f |)  P_2(| f |) } \begin{displaystyle} \sum_{\ell = |m|}^\infty p^*_{\ell m} \gamma^*_{\ell m}(f) \end{displaystyle}  \right) \right|^2 \over (\tilde Q, \tilde Q) }.
\ee
The optimal choice of filter function $Q(f)$ for determining the
$m$'th harmonic is the one which maximizes this ratio.  The largest
value is obtained by choosing
\be
\tilde Q_m(f) = 
{H(f)  \over 
P_1 (| f |)  P_2(| f |) } \begin{displaystyle} \sum_{\ell = |m|}^\infty p^*_{\ell m} \gamma^*_{\ell m}(f) \end{displaystyle} .
\ee
Using the definition (\ref{e:defh}) of $H(f)$ in terms of the spectral function
$\Omega_{\rm gw}(f)$, and substituting the optimal filter $\tilde Q_m$ into
the expression for $S/N$ yields
\be
\left( {S \over N} \right)_m^2 =
{9 H_0^4 \over 50 \pi^4} T \int_0^\infty df \>
{ \Omega_{\rm gw}^2(f) \over
f^6 P_1 ( f )  P_2( f ) } \left| \sum_{\ell = |m|}^\infty p_{\ell m}
\gamma_{\ell m}(f) \right|^2
 .
\ee
For any given source of stochastic gravitational waves, one can use
this formula for $(S/N)_m$ to determine the observation time $T$
required to observe the $m$'th harmonic of the signal as the earth
rotates relative to the cosmological frame.

In precise analogy with the analysis given in \cite{myreview}, the $m$'th
harmonic is observable with $90 \%$ confidence if $(S/N)_m$ exceeds $1.65$.

\section{Example: Dipole Induced by Proper Motion}
\label{s:ani}
It is well known that the electromagnetic background radiation,
generally referred to as the Cosmic Microwave Background Radiation
(CMBR), is highly isotropic.  The largest deviation from isotropy
results from the motion of our local system (the solar system
barycenter) with respect to the cosmological rest frame.  Analysis of
data from the Cosmic Background Explorer (COBE) satellite shows that
our local system is moving with a velocity $\beta_{\rm proper} \equiv
v/c = 0.001236$ in the direction $(l=264^\circ,b=48^\circ)$ in galactic
coordinates, or equivalently $(\alpha=168^\circ ,\delta=-7^\circ)$ in
celestial coordinates.  To lowest order in our proper velocity, this
gives rise to an anisotropy in the CMBR described by the temperature
distribution
\be
T(\gamma) = T_0 (1 + \beta_{\rm proper} \cos \gamma),
\ee
where $\gamma$ is the angle between a point on the sky and the velocity
vector of our local system, and $T_0=2.73 \rm K$ is the mean
temperature of the CMBR.

In this section, we address the question: ``Would a corresponding
dipole moment in the stochastic gravity wave background, arising from
the proper motion of the solar system barycenter, be observable with
either the initial or advanced LIGO detectors?"
In the previous section, we calculated the Signal-to-Noise ratio for
observations of the $m$'th harmonic $S_m$ (with respect to the earth's
rotation) of the signal (obtained by correlating two gravitational-wave
detectors).  This is determined entirely by the quantities
\beau
H_0 & = & {\rm Hubble\ expansion\ rate,\ sec}^{-1}. \\
T & = & {\rm Total\ observation\ time,\ sec}. \\
\gamma_{\ell m}(f) & = & {\rm
Overlap\ reduction\ functions,\ dimensionless}.\\
\Omega_{\rm gw} & = & {\rm
Fractional\ energy\ density\ in\ gravitational\ waves,\ dimensionless}.\\
p_{\ell m} & = & {\rm Normalized\ multipoles\ of\ gw\ stochastic\ background,\ dimensionless}.\\
P_i & = & {\rm Noise\ power\ spectral\ density\ of\ detector\ {\it i},\ sec}.
\eeau
As an example, we use this formula to answer the question posed above.

Let us make the following reasonable assumptions:\\
(1) $\Omega_{\rm gw}(f)$ is constant in the LIGO band, and\\
(2) the stochastic gravitational-wave background is isotropic in the same rest frame
    as the CMBR.\\
The anisotropies in the stochastic background resulting from our proper
motion are then described by
\be
\sum_{\ell m} p_{\ell m} Y_{\ell m} (\bar \theta,\bar \phi)  = 1 +
\beta_{\rm proper} (\cos \bar \theta \cos 97^\circ+ 
\sin \bar \theta \sin 97^\circ \cos \bar \phi)
\ee
where the angles above are standard spherical coordinates in the
(barred) earth frame.  One thus obtains the following multipole moments
\bea
p_{00} &=& \sqrt{4 \pi} \cr
p_{1,-1} &=& \beta_{\rm proper} \sqrt{2 \pi \over 3} \sin 97^\circ \cr
p_{1,0} &=& \beta_{\rm proper} \sqrt{4 \pi \over 3} \cos 97^\circ \cr
p_{1,1} &=& -\beta_{\rm proper} \sqrt{2 \pi \over 3} \sin 97^\circ \cr
p_{\ell m} &= &0 \quad {\rm for\ } \ell>1.
\eea
To detect this signal, the optimal filter functions are
\be
Q_m(f) = C_m {1  \over
   f^3 P_1(|f|) P_2(|f|) } \sum_{\ell = |m|}^\infty p^*_{\ell m}\gamma^*_{\ell m} ,
\ee
where $C_m$ is an (irrelevant) normalization constant.  Making use of
the optimal filter functions $Q_0$ and $Q_1$ for the monopole and
dipole terms, we can make predictions about how large $\Omega_{\rm gw}$
needs to be in order that $S_0$ and $S_1$ are observable with a given
level of confidence in a given observation time.

We can express the sensitivity of a search for the $m$'th harmonic in
terms of the minimum value of $\Omega_{\rm gw}$ necessary to observe it
with 90\% confidence.  For $90\%$ confidence we need a Signal-to-Noise
of $1.65$.  The minimum value of $\Omega_{\rm gw}$ is then given for
the $m$'th harmonic by
\be
\Omega_{\rm 90\%}^{(m)} = (1.65) \sqrt{9 \over 50} {H_0^2 \over \pi^2}
T \left[
\int_0^\infty { df \>  \over
f^6 P_1 ( f )  P_2( f ) }
 \left| \sum_{\ell = |m|}^\infty p_{\ell m}
\gamma_{\ell m}(f) \right|^2
 \right]^{-{1 \over 2}}.
\label{e:omega90}
\ee
These values are shown in Table~\ref{t:sens} for the initial and
advanced LIGO detectors.  We note that there are potential sources of
sufficient intensity that a dipole might be observable with the
advanced LIGO detector.  These include stochastic backgrounds due to
cosmic strings or to a population of unresolved cosmological-distance
supernovae.

\begin{table}
\caption{Sensitivity of Initial and Advanced LIGO detectors to a dipole term
in the gravitational stochastic background, arising from motion of our local
system.  This table shows the intensity of stochastic background required to
detect either the monopole ($S_0$) or dipole ($S_1$) term in the signal, with
90\% confidence, in one year of observation.}
\begin{tabular}{ccc}
 & Initial LIGO & Advanced LIGO \\ \tableline \\
Monopole (m=0) 
    & $ \Omega_{90\%} = 1.6 \times 10^{-6} h^{-2}_{100} $
    & $ \Omega_{90\%} = 1.6 \times 10^{-11} h^{-2}_{100} $ \\
Dipole (m=1) 
    & $ \Omega_{90\%} =  2.5 \times 10^{-3} h^{-2}_{100} $
    & $ \Omega_{90\%} =  5.3 \times 10^{-8} h^{-2}_{100}  $ \\
\end{tabular}
\label{t:sens}
\end{table}

\section{Filter Normalizations and Signal Strengths}
\label{s:norm}
Specific models for an anisotropic background predict signal harmonics
$S_m$ of definite amplitudes and phases.  The phase of a given $S_m$
depends upon the chosen origin of time; shifting the origin of time
changes the phase of $S_m$ but not its amplitude. In order to falsify
or confirm a particular anisotropic model, one needs to predict the set
of complex numbers $S_m$.  However these numbers depend upon the
normalization $C_m$ of the optimal filter function $Q_m(f)$ so in order to
make predictions, we need to adopt a convention for these.

We have already chosen optimal filters for which the expected value of
$S_m$ is real.  At least upon first inspection, there do not appear to
be any very convenient choices for the overall scale of normalization.
Here, for illustrative purposes we adopt (what appears to us as) the
least disagreeable of these.  {\it We choose the normalization $C_m$ so
that the expected amplitude of the noise${}^2$ is unity:} $\langle
|N_m|^2 \rangle=1$.  We also assume that $C_m$ is positive and real.
Note that with this choice, the ``signal" is dimensionless.  From
equation (\ref{e:meannm}) this gives
\be
C_m = {2 \sqrt{T} \over \tau} \left[  \int_{-\infty}^{\infty} {df
\over  P_1 (| f |)  P_2(| f |) } \left| \sum_{\ell=|m|}^\infty p_{\ell
m} \gamma_{\ell m}(f) \right|^2
\right]^{-{1 \over 2}}.
\ee
Having chosen a normalization, we can list the expected signals for a given
detector.  If the actual signal value exceeds $1.65$ then it has been
detected with $90\%$ confidence.
For example, in the case of the dipole source just discussed,
the expected signals are given by (\ref{e:meansm}) as
\be
\langle S_m \rangle =
{3 H_0^2 \over 10 \pi^2} \sqrt{T} \Omega_{\rm gw}
\left[  \int_{-\infty}^{\infty} {df
\over  P_1 (| f |)  P_2(| f |) } \left| \sum_{\ell=|m|}^\infty p_{\ell
m} \gamma_{\ell m}(f) \right|^2
\right]^{{1 \over 2}}.
\ee
Evaluating this for our previous example gives:
\bea
\langle S_0 \rangle &=& \cases{\displaystyle 
1.65 \sqrt{T \over 1 {\rm \ year}} {\Omega_{\rm gw} h_{100}^2 \over 1.6 \times 10^{-6}}
 & for initial LIGO \cr
\displaystyle 1.65 \sqrt{T \over 1 {\rm \ year}} {\Omega_{\rm gw} h_{100}^2 \over 1.6 \times 10^{-11}}
 & for advanced LIGO
} \cr & & \cr
\langle S_1 \rangle &=& \cases{ 
\displaystyle 1.65 \sqrt{T \over 1 {\rm \ year}} {\Omega_{\rm gw} h_{100}^2 \over 2.5 \times 10^{-3}}
 & for initial LIGO \cr
\displaystyle 1.65 \sqrt{T \over 1 {\rm \ year}} {\Omega_{\rm gw} h_{100}^2 \over 5.3 \times 10^{-8}}
 & for advanced LIGO
} \cr & & \cr
\langle S_m \rangle &=& 0 {\rm \ for\ } m \ge 2 
\eea

\section{Galactic Sources}
\label{s:galactic}
In this section we consider the possibility of detecting anisotropies
in the stochastic gravitational wave background assuming that this
background either originates in,  or is scattered in the
same way as, the luminous matter in our galaxy, and for
sources distributed
in the same way as the galactic halo.  It appears very unlikely that in the
LIGO/VIRGO/GEO frequency band there
are any sources of a stochastic gravitational-wave background
distributed in  this way.  However the absence of such harmonics would
be one way to demonstrate that a
stochastic background had extra-galactic origin. 

For the distribution of luminous matter in the galaxy we consider a set
of three models constructed by Kent, Dame and Fazio\cite{KDF} to model
 $2{\cdot}4\,\mu$m data from the Infrared Telescope taken as part of
the {\sl Spacelab 2} mission.  These models all assume cylindrical
symmetry and take the total luminosity to consist of two
components, one, $\nu_D(r,z)$, modeling the disk and the other,
 $\nu_B(r,z)$, modeling the central bulge, where $r$ and $z$ denote
cylindrical polar coordinates based at the center of the Galaxy.

The first model, which we refer to as KDF1, takes
\bea
\nu_D(r,z) &=& \mu_D e^{-r/h_r} {\sech^2\bigl(z/(2 h_z)\bigr) \over 4
h_z}\\
\nu_B(r,z) &=& \mu_B {K_0\Bigl(\bigl[r^4+\bigl(z/(1-\epsilon_B)\bigr)^4\bigr]^{1/4}
     /h_B \Bigr) \over \pi h_B}
\eea
where $\mu_D =1072\,L_{\odot}$pc${}^{-2}$, $h_r=2775\,$pc, $h_z=121\,$pc,  
$\mu_B =6208\,L_\odot$pc${}^{-2}$, $h_B=634\,$pc and  $\epsilon_B=0.26$.
Here $K_0$ denotes the modified Bessel function or order 0.

The second model (KDF2) takes
\bea
\nu_D(r,z) &=& \mu_D e^{-r/h_r} {\exp\Bigl(-\bigl| z/(2 h_z)\bigr| \Bigr) \over 2
h_z}\\
\nu_B(r,z) &=& \mu_B {K_0\Bigl(\bigl[r^4+\bigl(z/(1-\epsilon_B)\bigr)^4\bigr]^{1/4}
     /h_B \Bigr) \over \pi h_B}
\eea
where $\mu_D =1208\,L_{\odot}$pc${}^{-2}$, $h_r=2694\,$pc, $h_z=204\,$pc,  
$\mu_B =7710\,L_\odot$pc${}^{-2}$, $h_B=500\,$pc and
$\epsilon_B=0.19$.

The third model (KDF3) takes
\bea
\nu_D(r,z) &=& \mu_D e^{-r/h_r} {\exp\Bigl(-\bigl| z/(2 h_z(r))\bigr| \Bigr) \over 2
h_z(r)}\\
\nu_B(r,z) &=& \mu_B {K_0\Bigl(\bigl[r^4+\bigl(z/(1-\epsilon_B)\bigr)^4\bigr]^{1/4}
     /h_B \Bigr) \over \pi h_B}
\eea
where $\mu_D =978\,L_{\odot}$pc${}^{-2}$, $h_r=3001\,$pc,   
$\mu_B =7395\,L_\odot$pc${}^{-2}$, $h_B=667\,$pc,
$\epsilon_B=0.39$ and $h_z(r)$ is now a function given by
\be
   h_z(r) = \cases{ h_{\rm min} &$r<r_{\rm min}$ \cr &\cr
                  \displaystyle h_{\rm min} + (h_\odot -  h_{\rm min}) 
                 {(r-r_{\rm min}) \over (r_\odot - r_{\rm min})}
                   & $r\geq r_{\rm min}$}
\ee
with $h_{\rm min} = 165\,$pc,  $r_{\rm min}=5300\,$pc, $h_\odot =
247\,$pc and $r_\odot = 8000\,$pc, the last two quantities being the
height of the galactic disk at the position of the Sun and the
distance of the Sun from the Galactic center, respectively.

Finally, as a model for the galactic halo we take the model of
Young \cite{young} 
\be
    \nu_{\rm halo} = \mu_{\rm halo} {\exp\left[ -7.669(R/R_e)^{1/4}
\right]    \over (R/R_e)^{7/8}}    \qquad R>0.2R_e
\ee
where $R = (r^2+z^2)^{1/2}$ denotes the distance from the Galactic center and
$R_e=2700\,$pc. 
\newpage

The final piece of information we need before we can calculate the
multipole moments $p_{\ell m}$ for such a distribution is the direction 
to the Galactic center and the orientation of the Galaxy.  
It is standard to express this in terms of the equatorial coordinate
system \cite{mihalas} in which the $z$-axis is taken along the Celestial
North pole, the $x$-axis is taken in the direction of the (1950)
vernal equinox. $90^\circ-\delta$ and $\alpha$ are taken as the spherical polar
coordinates corresponding to these axes.   In terms of these,
the direction to the Galactic center is given by 
\be
\alpha = 265.6^\circ \qquad \delta = -28.9^\circ  ,
\ee
and the direction of the Galactic North pole by
\be
\alpha = 192.25^\circ \qquad \delta = 27.4^\circ  .
\ee

We take the equatorial coordinate system to define our Cosmic frame. 
We now determine $P(\hat \Omega)$ by choosing a direction $\hat
\Omega$ and integrating the intensity along that
direction. 
The number of sources in an  element of solid angle $d\Omega$ at distance
is $D$ from the Earth is  proportional to  $\nu \;D^2 dD d\Omega$.
The intensity of the source drops like $1/D^2$ and hence the
integrated intensity 
is just proportional to $\int \nu dD$.
 In this way we arrive at $p_{\ell m}$   for each model
and hence by Eq.~(\ref{e:omega90}) the minimum value of $\Omega_{\rm gw}$
necessary for detection of each multipole.  To ensure convergence of
the sum in this case it was necessary to include contributions to the sum
for $\ell$ up to of
order 100 (dependent on $m$). 
Tables \ref{t:gal_sens_initial} and \ref{t:gal_sens_adv} show 
the intensity of the stochastic 
background distributed in this way required to
detect multipole moments from $m=0$ to $m=24$ for the four models
with 90\% confidence in one year of observation for Initial and
Advanced LIGO respectively.

\vfill

\begin{table}
\caption{Sensitivity of the Initial  LIGO detector to the
first 25 multipoles in the gravitational stochastic background,
assumed to follow the luminosity of the Galaxy or Galactic halo.
 This table shows the intensity of stochastic background required to
detect the multipole  $S_m$ with
90\% confidence in one year of observation.}
\begin{tabular}{ccccc}
m &$h^2_{100} \Omega_{90\%}$ KDF1&$h^2_{100} \Omega_{90\%}$ KDF2&$h^2_{100} \Omega_{90\%}$ KDF3&$h^2_{100} \Omega_{90\%}$ Halo \\ 
\tableline \\

 0   & $ 1.4 \times 10^{-5} $
    & $ 1.4 \times 10^{-5} $
     & $ 1.4 \times 10^{-5} $
    & $ 3.9 \times 10^{-3} $ \\
1 
    & $  1.4 \times 10^{-5} $
    & $  1.3 \times 10^{-5}  $ 
   & $ 1.3 \times 10^{-5} $
    & $ 3.4 \times 10^{-3} $ \\

2   & $ 1.9 \times 10^{-5} $
    & $ 1.9 \times 10^{-5} $
     & $ 1.8 \times 10^{-5} $
    & $ 3.6 \times 10^{-3} $ \\

3   & $ 3.0 \times 10^{-5} $
    & $ 2.9 \times 10^{-5} $
     & $ 2.9 \times 10^{-5} $
    & $ 4.2 \times 10^{-3} $ \\

4   & $ 4.6 \times 10^{-5} $
    & $ 4.4 \times 10^{-5} $
     & $ 4.6 \times 10^{-5} $
    & $ 5.0 \times 10^{-3} $ \\

5   & $ 7.3 \times 10^{-5} $
    & $ 6.9 \times 10^{-5} $
     & $ 7.1 \times 10^{-5} $
    & $ 6.1 \times 10^{-3} $ \\

 6   & $ 7.8 \times 10^{-5} $
    & $ 7.4 \times 10^{-5} $
     & $ 7.5 \times 10^{-5} $
    & $ 8.1 \times 10^{-3} $ \\
 
 7   & $ 8.6 \times 10^{-5} $
    & $ 8.0 \times 10^{-5} $
     & $ 8.3 \times 10^{-5} $
    & $ 1.2 \times 10^{-2} $ \\ 

 8   & $ 1.2 \times 10^{-4} $
    & $ 1.1 \times 10^{-4} $
     & $ 1.1 \times 10^{-4} $
    & $ 1.7 \times 10^{-2} $ \\

 9   & $ 2.0 \times 10^{-4} $
    & $ 2.0 \times 10^{-4} $
     & $ 1.9 \times 10^{-4} $
    & $ 2.7 \times 10^{-2} $ \\
10 
    & $  3.6 \times 10^{-4} $
    & $  4.0 \times 10^{-4}  $ 
   & $ 3.5 \times 10^{-4} $
    & $ 4.3 \times 10^{-2} $ \\

11   & $ 5.0 \times 10^{-4} $
    & $ 5.2 \times 10^{-4} $
     & $ 4.9 \times 10^{-4} $
    & $ 7.0 \times 10^{-2} $ \\

12   & $ 6.3 \times 10^{-4} $
    & $ 6.2 \times 10^{-4} $
     & $ 6.2 \times 10^{-4} $
    & $ 1.2 \times 10^{-1} $ \\

13   & $ 9.2 \times 10^{-4} $
    & $ 8.9 \times 10^{-4} $
     & $ 9.0 \times 10^{-4} $
    & $ 2.0 \times 10^{-1} $ \\

14   & $ 1.5 \times 10^{-3} $
    & $ 1.5 \times 10^{-3} $
     & $ 1.5 \times 10^{-3} $
    & $ 3.1 \times 10^{-1} $ \\

15   & $ 2.4 \times 10^{-3} $
    & $ 2.5 \times 10^{-3} $
     & $ 2.4 \times 10^{-3} $
    & $ 5.0 \times 10^{-1} $ \\
 
16   & $ 3.6 \times 10^{-3} $
    & $ 3.7 \times 10^{-3} $
     & $ 3.6 \times 10^{-3} $
    & $ 7.8 \times 10^{-1} $ \\ 

17   & $ 5.1 \times 10^{-3} $
    & $ 5.2 \times 10^{-3} $
     & $ 5.1 \times 10^{-3} $
    & $ 1.2 \times 10^{0} $ \\
 
18   & $ 7.3 \times 10^{-3} $
    & $ 7.2 \times 10^{-3} $
     & $ 7.1 \times 10^{-3} $
    & $ 1.7 \times 10^{0} $ \\

19 
    & $  1.0 \times 10^{-2} $
    & $  1.0 \times 10^{-2}  $ 
   & $ 1.0 \times 10^{-2} $
    & $ 2.4 \times 10^{0} $ \\

20   & $ 1.4 \times 10^{-2} $
    & $ 1.4 \times 10^{-2} $
     & $ 1.4 \times 10^{-2} $
    & $ 3.6 \times 10^{0} $ \\

21   & $ 2.1 \times 10^{-2} $
    & $ 2.1 \times 10^{-2} $
     & $ 2.0 \times 10^{-2} $
    & $ 5.0 \times 10^{0} $ \\

22   & $ 2.9 \times 10^{-2} $
    & $ 2.9 \times 10^{-2} $
     & $ 2.8 \times 10^{-2} $
    & $ 6.5 \times 10^{0} $ \\

23   & $ 4.0 \times 10^{-2} $
    & $ 3.9 \times 10^{-2} $
     & $ 3.8 \times 10^{-2} $
    & $ 8.9 \times 10^{0} $ \\

24   & $ 5.1 \times 10^{-2} $
    & $ 5.0 \times 10^{-2} $
     & $ 4.9 \times 10^{-2} $
    & $ 1.2 \times 10^{1} $ \\

\end{tabular}
\label{t:gal_sens_initial}
\end{table}

\newpage

\vbox{

\begin{table}
\caption{Sensitivity of the  Advanced LIGO detector to the
first 25 multipoles in the gravitational stochastic background,
assumed to follow the luminosity of the Galaxy or Galactic halo.
 This table shows the intensity of stochastic background required to
detect the multipole  $S_m$ with
90\% confidence in one year of observation.}

\begin{tabular}{ccccc}
m &$h^2_{100} \Omega_{90\%}$ KDF1&$h^2_{100} \Omega_{90\%}$ KDF2&$h^2_{100} \Omega_{90\%}$ KDF3&$h^2_{100} \Omega_{90\%}$ Halo \\ 
\tableline \\
 
 0   & $ 1.8 \times 10^{-10} $
    & $ 1.7 \times 10^{-10} $
     & $ 1.7 \times 10^{-10} $
    & $ 6.7 \times 10^{-8} $ \\
1 
    & $  4.8 \times 10^{-10} $
    & $  4.7 \times 10^{-10}  $ 
   & $ 4.7 \times 10^{-10} $
    & $ 7.8 \times 10^{-8} $ \\

2   & $ 1.1 \times 10^{-9} $
    & $ 1.0 \times 10^{-9} $
     & $ 1.0 \times 10^{-9} $
    & $ 1.1 \times 10^{-7} $ \\

3   & $ 5.3 \times 10^{-9} $
    & $ 5.0 \times 10^{-9} $
     & $ 5.4 \times 10^{-9} $
    & $ 2.6 \times 10^{-7} $ \\

4   & $ 6.4 \times 10^{-9} $
    & $ 6.0 \times 10^{-9} $
     & $ 6.3 \times 10^{-9} $
    & $ 6.4 \times 10^{-7} $ \\

5   & $ 1.1 \times 10^{-8} $
    & $ 1.0 \times 10^{-8} $
     & $ 1.0 \times 10^{-8} $
    & $ 1.6 \times 10^{-6} $ \\

 6   & $ 3.1 \times 10^{-8} $
    & $ 3.0 \times 10^{-8} $
     & $ 3.0 \times 10^{-8} $
    & $ 4.4 \times 10^{-6} $ \\
 
 7   & $ 1.2 \times 10^{-7} $
    & $ 1.3 \times 10^{-7} $
     & $ 1.2 \times 10^{-7} $
    & $ 1.3 \times 10^{-5} $ \\ 

 8   & $ 2.4 \times 10^{-7} $
    & $ 2.4 \times 10^{-7} $
     & $ 2.4 \times 10^{-7} $
    & $ 2.7 \times 10^{-5} $ \\

 9   & $ 4.5 \times 10^{-7} $
    & $ 4.2 \times 10^{-7} $
     & $ 4.5 \times 10^{-7} $
    & $ 6.3 \times 10^{-5} $ \\
10 
    & $  1.1 \times 10^{-6} $
    & $  1.0 \times 10^{-6}  $ 
   & $ 1.1 \times 10^{-6} $
    & $ 1.3 \times 10^{-4} $ \\

11   & $ 3.0 \times 10^{-6} $
    & $ 3.3 \times 10^{-6} $
     & $ 3.1 \times 10^{-6} $
    & $ 2.6 \times 10^{-4} $ \\

12   & $ 5.5 \times 10^{-6} $
    & $ 5.5 \times 10^{-6} $
     & $ 5.7 \times 10^{-6} $
    & $4.5 \times 10^{-4} $ \\

13   & $ 8.3 \times 10^{-6} $
    & $ 7.7 \times 10^{-6} $
     & $ 8.7 \times 10^{-6} $
    & $ 7.7 \times 10^{-4} $ \\

14   & $ 1.5 \times 10^{-5} $
    & $ 1.5 \times 10^{-5} $
     & $ 1.6 \times 10^{-5} $
    & $ 1.3 \times 10^{-3} $ \\

15   & $ 3.0 \times 10^{-5} $
    & $ 3.2 \times 10^{-5} $
     & $ 3.1 \times 10^{-5} $
    & $ 2.0 \times 10^{-3} $ \\
 
16   & $ 4.4 \times 10^{-5} $
    & $ 4.4 \times 10^{-5} $
     & $ 4.7 \times 10^{-5} $
    & $ 3.0 \times 10^{-3} $ \\ 

17   & $ 6.0 \times 10^{-5} $
    & $ 5.6 \times 10^{-5} $
     & $ 6.4 \times 10^{-5} $
    & $ 4.4 \times 10^{-3} $ \\
 
18   & $ 9.4 \times 10^{-5} $
    & $ 9.0 \times 10^{-5} $
     & $ 1.0 \times 10^{-4} $
    & $ 6.4 \times 10^{-3} $ \\

19 
    & $  1.6 \times 10^{-4} $
    & $  1.6 \times 10^{-4}  $ 
   & $ 1.7 \times 10^{-4} $
    & $ 9.0 \times 10^{-3} $ \\

20   & $ 2.1 \times 10^{-4} $
    & $ 2.1 \times 10^{-4} $
     & $ 2.3 \times 10^{-4} $
    & $ 1.2 \times 10^{-2} $ \\

21   & $ 2.7 \times 10^{-4} $
    & $ 2.5 \times 10^{-4} $
     & $ 3.0 \times 10^{-4} $
    & $ 1.7 \times 10^{-2} $ \\

22   & $ 3.9 \times 10^{-4} $
    & $ 3.7 \times 10^{-4} $
     & $ 4.2 \times 10^{-4} $
    & $ 2.3 \times 10^{-2} $ \\

23   & $ 5.9 \times 10^{-4} $
    & $ 5.8 \times 10^{-4} $
     & $ 6.4 \times 10^{-4} $
    & $ 3.0 \times 10^{-2} $ \\

24   & $ 7.8 \times 10^{-4} $
    & $ 7.5 \times 10^{-4} $
     & $ 8.4 \times 10^{-4} $
    & $ 3.9 \times 10^{-2} $ \\

\end{tabular}
\label{t:gal_sens_adv}
\end{table}

\section{Conclusion}
In this paper, we have shown how the signals from a pair of
gravitational wave detectors may be analyzed to search for anisotropies
in the stochastic gravitational wave background.  We have shown how the
correlation between two detectors may be determined with an averaging
time short compared to a day but long compared to the light travel time
between the detectors, and how this correlation may be decomposed into
harmonics of the Earth's rotation.  We have calculated the
signal-to-noise ratios associated with such measurements, and shown
that certain types of anisotropy might reasonably be detected with
instruments that will be available in the not-too-distant future.  For
example the anisotropy due to our motion with respect to the
cosmological rest frame might well be observable with a
second-generation LIGO detector.

At this point, it is difficult to proceed further without a more
detailed knowledge of the instrumental data that will be forthcoming.
The results given in this paper make it straightforward to predict the
expected harmonic amplitudes $S_m$ of the detector correlation for a
given anisotropic distribution $p_{\ell m}$ of gravitational wave
background.  It is more difficult to go the other way.  This would
involve using a given set of observed harmonic amplitudes $S_m$ to
obtain the values of (or constraints on) the $p_{\ell m}$.  There are a
variety of fitting techniques that could be used -- making the
appropriate choice will probably require real data.

\acknowledgements
This work has been partially supported by the National Science
Foundation grant PHY95-07740, by the Forbairt grant SC/96/712, and by
the LIGO visitors program PHY92-10038.

}
\appendix
\section*{Analytic evaluation of integrals}

In this appendix we derive a closed form expression for the integral
 \be
{\cal I}^k_{\ell}(x) =  \int_{-1}^1  du \; {\rm e}^{i u x}
{\rm P}^k_\ell( u) u^N (1-u^2)^{|k|/2}
\ee
where $N$ is a  non-negative integer.

First we note that
\be 
 {\rm P}^{-k}_\ell( u) = (-1)^k {(l-k)! \over (l+k)!} {\rm P}^k_\ell(
 u)
\ee
so that it is only necessary to deal with the case $k \geq 0$.
For $k \geq 0$ we have
\be 
 {\rm P}^{k}_\ell( u) = (-1)^k (1-u^2)^{k/2} {d^k \over du^k} P_\ell(u)
\ee
giving 
 \be
{\cal I}^k_{\ell}(x)= (-1)^k \int_{-1}^1  du \; {\rm e}^{i u x}
 u^N (1-u^2)^{k} {d^k \over du^k} P_\ell(u) .
\ee
The presence of the factor $(1-u^2)^{k}$ now ensures that when
we integrate by parts $k$ times no boundary terms appear so that
 \be
{\cal I}^k_{\ell}(x)= \int_{-1}^1  du \; P_\ell(u) {d^k \over du^k}
\left[ {\rm e}^{i u x} u^N (1-u^2)^{k} \right] .
\ee
The derivative can be expanded by the Liebniz rule to give
\be 
  \sum\limits_{r=0}^k (ix)^r p_{N+k+r}(u) {\rm e}^{i u x}
\ee
where  $p_{N+k+r}(u)$ is a polynomial of degree $N+k+r$.
Our problem is thus reduced to that of finding
\be
  {\cal J}^M_{\ell}(x)  =  \int_{-1}^1  du \; {\rm e}^{i u x} u^M
  P_\ell(u)  .
\ee 
For $M=0$ this is an elementary integral given by
\be  
{\cal J}^0_{\ell}(x) = 2 i^\ell j_\ell(x)   \label{J0}
\ee 
where
$ j_\ell(x)$ denotes the spherical Bessel function of order $\ell$
\cite{Jackson}.
Higher $M$ values may then be obtained by differentiation
\be
    {\cal J}_{\ell }^M (x) = (-i)^M {d^M \over dx^M} {\cal J}_{{\ell}}^0(x)
    .
\ee
These derivatives may in turn be expressed back in terms of
(undifferentiated) spherical Bessel functions using the
relations\cite{Jackson}
\bea
   {d \over dx} j_\ell(x) &=& {1 \over 2 \ell + 1} 
 \left[\ell j_{\ell-1}(x) - (\ell + 1) j_{\ell+1}(x)\right]\\ 
  &=& {\ell \over x} j_\ell(x) - j_{\ell+1}(x) . 
\label{e:sphbess}
\eea
For $M=1$ to 8 we have
\beau
{\cal J}^1_\ell (x) &=& {\ell \over (2\ell + 1)}j_{\ell-1} (x) -
                      {(\ell + 1) \over (2\ell+ 1)} j_{\ell+1} (x) \\
{\cal J}^2_\ell (x) &=& {(\ell-1) \ell  \over (2\ell -1 )(2\ell+1)} j_{\ell -2}(x) - 
   { (2 \ell^2 + 2\ell -1) \over (2\ell-1)(2\ell+3)} j_\ell (x) + 
   {(\ell +1)(\ell + 2) \over (2\ell+1)(2\ell+3)} j_{\ell +2}   (x) \\
{\cal J}^3_\ell (x) &=&  {\ell(\ell-1)(\ell-2) \over (2\ell-3)(2\ell-1)(2\ell+1)}j_{\ell -3}(x) - 
   {3\ell(\ell^2-2) \over (2\ell-3)(2\ell+1)(2\ell+3)}j_{\ell-1}(x) + \\&&
   {3(\ell+1)(\ell^2+2\ell-1) \over (2\ell-1)(2\ell+1)(2\ell+5)}j_{\ell +1}(x) - 
   {(\ell+1)(\ell+2)(\ell+3) \over (2\ell+1)(2\ell+3)(2\ell+5)}j_{\ell+3}(x) \\
{\cal J}^4_\ell (x) &=&  { \ell(\ell-1)(\ell-2)(\ell-3)\over
       (2\ell+1)(2\ell-1)(2\ell-3)(2\ell-5)} j_{\ell -4}(x) -   
 {2\ell (\ell-1)(2\ell^2 -2\ell -7)) \over 
       (2\ell+3) (2\ell+1)(2\ell-1)(2\ell-5)} j_{\ell -2}(x)\\ && + 
 {3 (2\ell^4 + 4\ell^3 -6\ell^2 - 8\ell +3) \over
(2\ell+5)(2\ell+3)(2\ell-1)(2\ell-3)} j_\ell(x)
   -   {2(\ell+1)(\ell+2)(2\ell^2 +6\ell - 3) \over
(2\ell+7)(2\ell+3)(2\ell+1)(2\ell-1)} j_{\ell +2}(x) +\\
&&  {(\ell+1)(\ell+2)(\ell+3)(\ell+4) \over
(2\ell+7)(2\ell+5)(2\ell+3)(2\ell+1)} j_{\ell +4}(x) \\
{\cal J}^5_\ell (x) &=&  {\ell(\ell-1)(\ell-2)(\ell-3)(\ell-4) \over
   (2\ell-7)(2\ell-5)(2\ell-3)(2\ell-1)(2\ell+1)}j_{\ell -5}(x) -
   {5\ell(\ell-1)(\ell-2)(\ell^2 -2\ell - 5) \over
(2\ell-7)(2\ell-3)(2\ell-1)(2\ell+1)(2\ell+3)}j_{\ell -3}(x) +\\ &&
   {5\ell(2\ell^4-16\ell^2+23)\over (2\ell-5)(2\ell-3)(2\ell+1)(2\ell+3)(2\ell+5)}j_{\ell-1}(x) -
  {5(\ell+1)(2\ell^4+8\ell^3-4\ell^2-24\ell+9)  \over 
             (2\ell-3)(2\ell-1)(2\ell+1)(2\ell+5)(2\ell+7)} j_{\ell +1}(x) +\\&&
  {5(\ell+1)(\ell+2)(\ell+3)(\ell^2+4\ell-2) \over 
                    (2\ell-1)(2\ell+1)(2\ell+3)(2\ell+5)(2\ell+9)}j_{\ell+3}(x) -
 {(\ell+1)(\ell+2)(\ell+3)(\ell+4)(\ell+5) \over 
             (2\ell+1)(2\ell+3)(2\ell+5)(2\ell+7)(2\ell+9)}  j_{\ell +5}(x) \\
{\cal J}^6_\ell (x) &=& {\ell(\ell-1)(\ell-2)(\ell-3)(\ell-4)(\ell-5) \over 
                 (2\ell-9)(2\ell-7)(2\ell-5)(2\ell-3)(2\ell-1)(2\ell+1)}j_{\ell -6}(x) -\\ &&
  {3\ell(\ell-1)(\ell-2)(\ell-3)(2\ell^2-6\ell-13) \over (2\ell-9)(2\ell-5)(2\ell-3)(2\ell-1)(2\ell+1)(2\ell+3)}j_{\ell -4}(x) +\\&&
 {15\ell(\ell-1)(\ell^4-2\ell^3 -11 \ell^2 +12 \ell + 29) \over (2\ell-7)(2\ell-5)(2\ell-1)(2\ell+1)(2\ell+3)(2\ell+5)}
     j_{\ell -2}(x) -\\ &&
  {5(4\ell^6 + 12 \ell^5 -38 \ell^4 -96 \ell^3 + 88\ell^2 +138 \ell -45) \over
   (2\ell-5)(2\ell-3)(2\ell-1)(2\ell+3)(2\ell+5)(2\ell+7)}j_\ell (x) + \\&&
   {15(\ell+1)(\ell+2)(\ell^4 +6\ell^3 + \ell^2 -24 \ell + 9) \over (2\ell-3)(2\ell-1)(2\ell+1)(2\ell+3)(2\ell+7)(2\ell+9)}
      j_{\ell +2}(x) -\\&&
   {3(\ell+1)(\ell+2)(\ell+3)(\ell+4) (2\ell^2 +10\ell -5) \over
(2\ell-1)(2\ell+1)(2\ell+3)(2\ell+5)(2\ell+7)(2\ell+11)}j_{\ell +4}(x)
+ \\&&
  {(\ell+1)(\ell+2)(\ell+3)(\ell+4)(\ell+5)(\ell+6) \over (2\ell+1)(2\ell+3)(2\ell+5)(2\ell+7)(2\ell+9)(2\ell+11)}j_{\ell +6}(x)\\
{\cal J}^7_\ell (x) &=& {\ell(\ell-1)(\ell-2)(\ell-3)(\ell-4)(\ell-5)(\ell-6) \over
 (2\ell-11)(2\ell-9)(2\ell-7)(2\ell-5)(2\ell-3)(2\ell-1)(2\ell+1)} j_{\ell -7}(x) - \\&&
   {7\ell(\ell-1)(\ell-2)(\ell-3)(\ell-4) (\ell^2 -4\ell -8) \over 
         (2\ell-11)(2\ell-7)(2\ell-5)(2\ell-3)(2\ell-1)(2\ell+1)(2\ell+3)}j_{\ell -5}(x)  +\\&&
  {21\ell(\ell-1)(\ell-2) (\ell^4 -4\ell^3 -12\ell^2 +32\ell +54) \over (2\ell-9)(2\ell-7)(2\ell-3)(2\ell-1)(2\ell+1)(2\ell+3)(2\ell+5)}
     j_{\ell -3}(x) -\\&&
     {35\ell(\ell^2-6)(\ell^4 -14 \ell^2 +22) \over (2\ell-7)(2\ell-5)(2\ell-3)(2\ell+1)(2\ell+3)(2\ell+5)(2\ell+7)} j_{\ell-1}(x) +\\&&
   {35(\ell+1)(\ell^2 + 2\ell -5)*(\ell^4 + 4\ell^3 -8\ell^2 -24\ell+ 9) \over (2\ell-5)(2\ell-3)(2\ell-1)(2\ell+1)(2\ell+5)(2\ell+7)(2\ell+9)}
   j_{\ell +1}(x) - \\&&
 {21(\ell+1)(\ell+2)(\ell+3)(\ell^4 +8\ell^3 +6\ell^2 -40 \ell +15) \over
(2\ell-3)(2\ell-1)(2\ell+1)(2\ell+3) (2\ell+5)(2\ell+9)(2\ell+11) }  j_{\ell+3}(x) +\\&&
 { 7(\ell+1)(\ell+2)(\ell+3) (\ell+4)(\ell+5)(\ell^2+6\ell-3) \over (2\ell-1)(2\ell+1)(2\ell+3)(2\ell+5)(2\ell+7)(2\ell+9)(2\ell+13)}
   j_{\ell +5}(x) -\\&&
 {(\ell+1)(\ell+2)(\ell+3)(\ell+4)(\ell+5)(\ell+6)(\ell+7) \over  (2\ell+1)(2\ell+3)(2\ell+5)(2\ell+7)(2\ell+9)(2\ell+11)(2\ell+13)}
 j_{\ell   +7}(x) \\
{\cal J}^8_\ell (x) &=&
  {\ell(\ell-1)(\ell-2)(\ell-3)(\ell-4)(\ell-5)(\ell-6)(\ell-7) \over 
(2\ell-13)(2\ell-11)(2\ell-9)(2\ell-7)(2\ell-5)(2\ell-3)(2\ell-1)(2\ell+1)} j_{\ell -8}(x) - \\&&
  {4\ell(\ell-1)(\ell-2)(\ell-3)(\ell-4)(\ell-5)(2\ell^2-10\ell-19) \over 
       (2\ell-13)(2\ell-9)(2\ell-7)(2\ell-5)(2\ell-3)(2\ell-1)(2\ell+1)(2\ell+3)}j_{\ell -6}(x) +\\&&
 {14\ell(\ell-1)(\ell-2)(\ell-3)(2\ell^4-12\ell^3-22\ell^2+120\ell +173) \over
   (2\ell-11)(2\ell-9)(2\ell-5)(2\ell-3)(2\ell-1)(2\ell+1)(2\ell+3)(2\ell+5)}j_{\ell -4}(x) - \\&&
  {28\ell(\ell-1) (2\ell^6 -6\ell^5 -49 \ell^4 +108 \ell^3 +371 \ell^2 -426\ell -873) \over 
         (2\ell-9)(2\ell-7)(2\ell-5)(2\ell-1)(2\ell+1)(2\ell+3)(2\ell+5)(2\ell+7)}j_{\ell -2}(x) +\\&&
 {35(2\ell^8 +8\ell^7 -44\ell^6 -160 \ell^5 +286 \ell^4 +848 \ell^3 - 604 \ell^2 -1056 \ell
+315) \over (2\ell-7)(2\ell-5)(2\ell-3)(2\ell-1)(2\ell+3)(2\ell+5)(2\ell+7)(2\ell+9)}j_{\ell}(x) - \\&&
 {28(\ell+1)(\ell+2)(2\ell^6 +18\ell^5 +11\ell^4 -204 \ell^3 -157 \ell^2 +690 \ell -225) \over
          (2\ell-5)(2\ell-3)(2\ell-1)(2\ell+1)(2\ell+3)(2\ell+7)(2\ell+9)(2\ell+11)}j_{\ell +2}(x) +\\&&
  {14(\ell+1)(\ell+2)(\ell+3) (\ell+4)(2\ell^4 +20\ell^3 +26 \ell^2 -120\ell +45) \over
(2\ell-3)(2\ell-1)(2\ell+1)(2\ell+3)(2\ell+5)(2\ell+7)(2\ell+11)(2\ell+13)}j_{\ell +4}(x) -\\&&
 {4(\ell+1)(\ell+2)(\ell+3)(\ell+4)(\ell+5)(\ell+6)(2\ell^2 + 14\ell -7) \over 
(2\ell-1)(2\ell+1)(2\ell+3)(2\ell+5)(2\ell+7)(2\ell+9)(2\ell+11)(2\ell+15)}j_{\ell +6}(x) +\\&&
  {(\ell+1)(\ell+2)(\ell+3)(\ell+4)(\ell+5)(\ell+6)(\ell+7)(\ell+8) \over 
 (2\ell+1)(2\ell+3)(2\ell+5)(2\ell+7)(2\ell+9)(2\ell+11)(2\ell+13)(2\ell+15)}
j_{\ell +8}(x)
\eeau

Alternative forms may by obtained by use of Eq.~(\ref{e:sphbess}).

\newpage
%

\end{document}